\documentclass[pra,letterpaper,aps,10pt,superscriptaddress,twocolumn,floatfix,showpacs]{revtex4-1}
\usepackage{graphicx}
\usepackage{amsmath}
\usepackage{amsfonts}
\usepackage{float}
\usepackage{amssymb}
\usepackage{epsfig}
\usepackage{epstopdf}
\DeclareGraphicsExtensions{.pdf,.eps,.png,.jpg,.mps}
\usepackage[pdftex]{color}
\usepackage{amsmath,graphicx,amssymb,braket,xcolor,subfigure,upgreek}
\usepackage[colorlinks, linkcolor=blue, citecolor=blue, urlcolor=blue, breaklinks=true]{hyperref}
\usepackage{microtype}
\usepackage{bbm}
\usepackage{color}
\usepackage{dsfont}

\newcommand{\calS}{\mathcal{S}}
\newcommand{\calA}{\mathcal{A}}

\begin{document}

\title{Cooperative subwavelength molecular quantum emitter arrays}
\author{R. Holzinger}
\affiliation{Institut f\"{u}r Theoretische Physik, Universit\"{a}t Innsbruck, Technikerstrasse 21a, A-6020 Innsbruck, Austria}
\author{S.~A. Oh}
\affiliation{Max Planck Institute for the Science of Light, Staudtstra{\ss}e 2,
D-91058 Erlangen, Germany}
\author{M. Reitz}
\affiliation{Max Planck Institute for the Science of Light, Staudtstra{\ss}e 2,
D-91058 Erlangen, Germany}
\affiliation{Department of Physics, Friedrich-Alexander-Universit\"{a}t Erlangen-N\"urnberg, Staudtstra{\ss}e 7,
D-91058 Erlangen, Germany}
\author{H. Ritsch}
\affiliation{Institut f\"{u}r Theoretische Physik, Universit\"{a}t Innsbruck, Technikerstrasse 21a, A-6020 Innsbruck, Austria}
\author{C. Genes}
\affiliation{Max Planck Institute for the Science of Light, Staudtstra{\ss}e 2,
D-91058 Erlangen, Germany}
\affiliation{Department of Physics, Friedrich-Alexander-Universit\"{a}t Erlangen-N\"urnberg, Staudtstra{\ss}e 7,
D-91058 Erlangen, Germany}
\date{\today}

\begin{abstract}
Dipole-coupled subwavelength quantum emitter arrays respond cooperatively to external light fields as they may host collective delocalized excitations (a form of excitons) with super- or subradiant character. Deeply subwavelength separations typically occur in molecular ensembles, where in addition to photon-electron interactions, electron-vibron couplings and vibrational relaxation processes play an important role. We provide analytical and numerical results on the modification of super- and subradiance in molecular rings of dipoles including excitations of the vibrational degrees of freedom. While vibrations are typically considered detrimental to coherent dynamics, we show that molecular dimers or rings can be operated as platforms for the preparation of long-lived dark superposition states aided by vibrational relaxation. In closed ring configurations, we extend previous predictions for the generation of coherent light from ideal quantum emitters to molecular emitters, quantifying the role of vibronic coupling onto the output intensity and coherence.
\end{abstract}

\pacs{42.50.Ar, 42.50.Lc, 42.72.-g}

\maketitle
\section{Introduction}
Structured subwavelength arrays of quantum emitters allow for the coherent hopping of excitations via the near-field coupling of neighbouring dipoles~\cite{brixner2017exciton, moreno2019subradiance, mattioni2021design,popp2021quantum}. In addition, they exhibit correlated spontaneous emission and support super- and subradiant collective modes, which can be exploited to control the interaction with impinging light. Possible applications range from the design of highly reflective quantum metasurfaces~\cite{rui2020asubradiant,perczel2017photonic,rubies2022photon} to the engineering of platforms showing robust transport of excitation in topological quantum optics~\cite{perczel2017topological,bettles2017topological} and of high-fidelity photon storage devices for quantum information processing~\cite{plankensteiner2015selective, asenjo2017exponential, guimond2019subradiant}. Moreover, quantum emitter rings have been proposed to act as coherent light sources on the nanoscale~\cite{holzinger2020nanoscale}. \\
\indent While subwavelength separations are not easily achieved in standard quantum optics setups, molecular aggregates (i.e., arrays of identical molecules) can feature deeply subwavelength separations on the nanometer scale, while retaining the electronic structure of the individual dipole transitions~\cite{saikin2013photonics}. They can be artificially synthesized in a wide variety of forms such as one dimensional chains or two dimensional films and can also be found in nature, in particular in the photosynthetic light-harvesting complexes of plants and bacteria~\cite{cogdell2006architecture,ishizaki2012quantum, mattioni2021design}. For example, long-lived electronic quantum coherence in a light-harvesting protein (the Fenna-Matthews-Olson complex) has been experimentally observed~\cite{brixner2005two} and theoretically tackled~\cite{ishizaki2012quantum}. The downside of such systems is the much more complex structure, which introduces coupling of electronic degrees of freedom with intra- and inter-molecular vibrations. While it is well established that the strong coherent near-field interactions give rise to delocalized exciton states (of the so-called Frenkel excitons) in molecular aggregates~\cite{devoe1964optical, scholes2006exciton}, the characterization of the accompanying collective dissipation is usually not fully taken into account in such systems as vibronic couplings and induced dephasing are considered to dominate the dynamics, especially at high temperatures. It is therefore interesting and timely to characterize cooperative dissipative effects in the presence of vibrations, a task which involves an extension of previously developed methods to describe electron-photon-phonon interactions on an individual dipole basis~\cite{reitz2019langevin, reitz2020molecule}. Moreover, the addition of localized gain, renders molecular emitter arrays as possible candidates for the realization of nanoscale coherent light sources as recently introduced for pure quantum emitters~\cite{holzinger2020nanoscale}. As an alternative (or addition) to strong near-field coupling, a modified material response of a molecular ensemble can also be obtained by collective strong coupling of the ensemble to a cavity which creates similar delocalized excitaton states among the molecules which are then hybridized with the cavity mode. The field of molecular polaritonics has recently emerged as a platform for observing strong modifications of material properties such as charge and energy transport or chemical reactivity ~\cite{ orgiu2015conductivity, zhong2017energy, hutchinson2012modifying, herrera2020molecular}. \\
\begin{figure}[t]
\includegraphics[width=0.9\columnwidth]{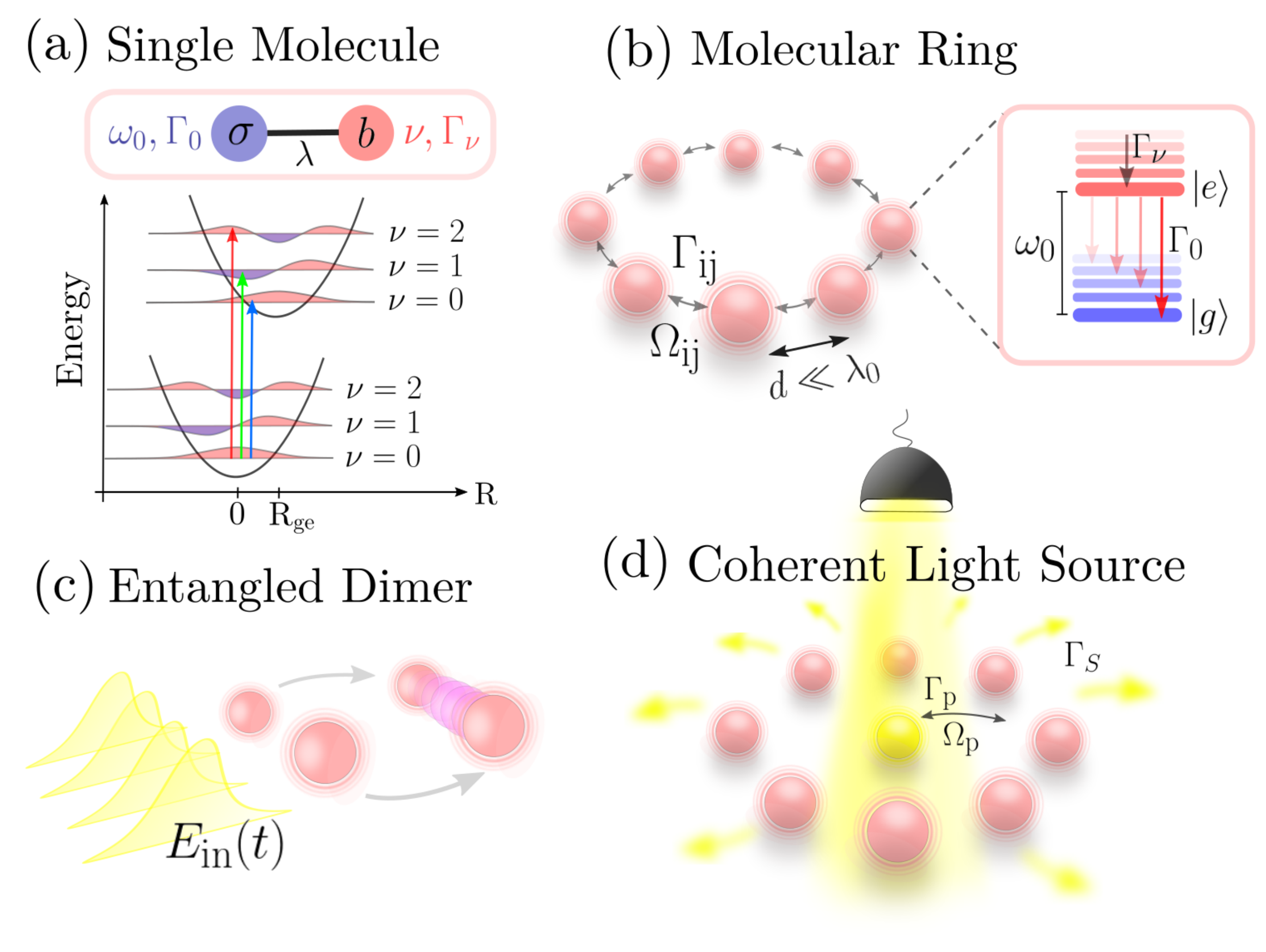}
\caption {(a) The equilibrium mismatch $R_\mathrm{ge}$ between the ground and excited state electronic potential landscapes along a given nuclear coordinate leads to the standard Franck-Condon physics with a branching of transitions into different vibrational levels. The electron-vibron coupling is schematically represented by the link, at coupling strength $\lambda$, between an electronic transition operator $\sigma$ and a bosonic vibrational mode operator $b$. (b) Schematics of a molecular ring where mutual interactions are mediated by the electromagnetic vacuum at coherent/incoherent rates $\Omega_\mathrm{ij}$ and $\Gamma_\mathrm{ij}$. The inset shows branching of electronic transitions between the manifolds of vibrational levels. (c) Preparation of an entangled molecular dimer with subwavelength separation $d\ll\uplambda_0$ via an impinging short laser pulse. (d) Schematics of a molecular nanoscale light source where the central gain molecule is incoherently pumped and coherently coupled to the symmetric eigenmode of the ring molecules. The ring provides an effective resonator enhancement leading to the emission of coherent laser light.}
\label{fig1}
\end{figure}
\indent In this work, we perform analytical and numerical studies of cooperative radiative properties of molecular arrays with particular emphasis on ring configurations, where we treat the vibronic coupling and electron-photon interactions on equal footing. Our treatment combines two approaches, a master equation approach, where the thermal environment of the vibrational degrees of freedom is traced out and a quantum Langevin equations approach, where the time evolution of both electronic and vibrational operators are fully considered. As a first important step, we elucidate the influence of vibronic couplings on the scaling of collective emission rates: modifications for the case of molecular systems originate from the Franck-Condon factors, which lead to a decay of the electronic coherence via coupling to several states of the vibrational degrees of freedom and the vibrational thermal environment.\\
\indent Analytical results can be derived and understood more easily by a transformation to a collective electronic basis, which involves a single bright (symmetric) state and many more dark (antisymmetric) state of typically superradiant and subradiant character respectively. This basis allows a simplified understanding of how standard scenarios, such as Dicke superradiance and the band structure of dipole-dipole induced transport of excitations, are modified by the electron-vibron interactions. \\
\indent While vibronic couplings are generally seen as detrimental in the efforts of controlling electronic coherence with light modes, here we present a generic vibronic dimer model, where bipartite long-lived entanglement is even engineered owing to vibrational relaxation. The system involves a nanometer spaced molecular dimer, where two chromophores exchange energy but not charge. Under favorable conditions, unidirectional flow of population for a driven symmetric collective state is directed into a robust, entangled antisymmetric state, via a process similar to the F\"orster resonance energy transfer occurring in acceptor/donor configurations.\\
\indent´The same transformation to a collective basis proves useful in the understanding of molecular nanorings illuminated by incoherent light sources, as recently proposed for the design of nanoscale coherent light sources~\cite{holzinger2020nanoscale}. In such systems, symmetric collective states are almost fully responsible for the generation of emitted light, which greatly aids our analytical and numerical analysis, allowing for a great reduction of the relevant Hilbert space dimension and therefore for numerical results with a reasonably sized molecular nano-rings, where each electronic transition is coupled to at least one own phonon mode.\\
\indent The paper is organized as follows: Sec.~\ref{sec:Model} introduces the open system dynamics formalism for molecules including electron-photon and electron-vibron interactions. In Sec.~\ref{Sec3}, we describe super- and subradiance both in the Dicke limit of closely spaced ensembles, for population inverted systems, and also in the weak excitation for arbitrarily spaced chains and rings. We then introduce in Sec.~\ref{Sec4} a particular case of nanoscale sized molecular dimers, where vibrationally induced couplings between collective symmetric and antisymmetric electronic states allow for the addressing of long-lived dark entangled states. The symmetric/antisymmetric collective basis is then generalized to the ring geometry with particular relevance to molecular nanoring lasers. In Sec.~\ref{Sec5}, we provide analytical and numerical results for the scaling of intensity and second order correlation functions of coherent light emitted by an incoherently pumped nanoscale molecular ring.

\section{Model}
\label{sec:Model}
We consider $\mathcal{N}$ identical molecular quantum emitters, each involving electronic transitions between two potential landscapes, with minima slightly shifted from each other along a nuclear coordinate. This mismatch of the electronic potential energy landscapes in the ground and excited states gives rise to the electron-vibron coupling, as depicted in Fig.~\ref{fig1}(a). External drive of electronic transitions is accompanied, in consequence, by the excitation of the motion of the nuclei, depicted as eigenstates of a harmonic potential in Fig.~\ref{fig1}(a). The electronic transition for molecule $j$ (index running between $1$ and $\mathcal{N}$) is at frequency splitting $\omega_0$ ($\hbar=1$) and described by the collapse operator $\sigma_j=\ket{g}_j\bra{e}_j$ and its Hermitian conjugate. The vibrational degree of freedom is at frequency $\nu$ and is described by a bosonic operator $b_j$ satisfying the commutation relations $\left[b_j,b_j^\dagger\right]=1$. The vibronic coupling is illustrated in Fig.~\ref{fig1}(a) as a link between the electronic and vibration operator with magnitude characterized by the Huang-Rhys factor $\lambda^2$. The electronic and vibrational degrees of freedom are subject to loss quantified by the spontaneous emission rate $\Gamma_0$ and by the vibrational relaxation rate  $\Gamma_\nu$, respectively. A standard Jablonski diagram of radiative and non-radiative processes involving two electronic states with their corresponding vibrational manifold is illustrated in the inset of Fig.~\ref{fig1}(b). This complex competition of transitions shows that molecules are typically inefficient quantum emitters as they do not possess closed transitions. Furthermore, we will consider rings of $\mathcal{N}$ molecules, as illustrated in Fig.~\ref{fig1}(b), with ring radius $r$ and interparticle separation $d=2r\sin{2\pi/\mathcal{N}}$. Their close separation brings into play cooperative effects such as near field dipole-dipole interactions and collective spontaneous emission, quantified by the distance dependent rates $\Omega_{ij}$ and $\Gamma_{ij}$, which are mediated by the quantum electromagnetic vacuum.\\
\indent The free Hamiltonian for the ensemble of $\mathcal{N}$ molecules $\mathcal{H}_0=\textstyle \sum_j h_0^{(j)}$ is obtained as a sum over each particle's free Hamiltonian
\begin{equation}
h_0^{(j)}=\left(\omega_0+\lambda^2 \nu\right)\sigma^\dagger_j\sigma_j+\nu b_{j}^\dagger b_{j},
\end{equation}
which sees a vibronic shift $\lambda^2 \nu$ added to the electronic bare transition frequency (which will later cancel out after a polaron transformation -- see Appendix~\ref{AppendixA}). The index $j$ runs from $1$ to $\mathcal{N}$ for the ring configuration which will be used in the next section to derive cooperative radiative emission properties of molecular ensembles. In Sec.~\ref{Sec5} we will incorporate an additional index $p$ to describe the situation depicted in Fig.~\ref{fig1}(d), which sees the realization of a molecular nanoscale light source with a gain molecule implanted in the center of the ring.\\
\indent The vibronic coupling Hamiltonian~\cite{holstein1959study} is now added as a sum $\mathcal{H}_{\text{Hol}}=\textstyle \sum_j h_{\text{Hol}}^{(j)}$ over all particles, where
\begin{equation}
\label{holstein}
h_{\text{Hol}}^{(j)}=-\lambda \nu \sigma_j^\dagger\sigma_j (b_{j}^\dagger+b_{j}).
\end{equation}
The Holstein Hamiltonian listed above assumes identical molecules and is a minimal model for electron-vibron interactions derivable from first principles~\cite{reitz2019langevin} (see Appendix~\ref{AppendixA}).\\
\indent For closely spaced quantum emitters, near-field dipole-dipole interactions at rates $\Omega_{jj'}$ are added, which are strongly dependent on their interseparation (with a standard $|\vec{r}_j-\vec{r}_{j'}|^{-3}$ dependence in the near field region) and relative orientation of transition dipoles~\cite{reitz2022cooperative} (see Appendix~\ref{AppendixB} for exact expressions). This can be listed as
\begin{equation}
\mathcal{H}_\text{d-d}=\textstyle \sum_{j\neq j'}\Omega_{jj'}\sigma^\dagger_j\sigma_{j'}
\label{Dipole-Ham}
\end{equation}
and describes an excitation transfer via a virtual photon exchange. Notice that by definition the diagonal terms $\Omega_{jj}$ vanish.\\
\indent To the coherent dynamics one can then add the effects of infinite reservoirs in an open system dynamics described by a master equation (for the system's density operator $\rho$) in the form
\begin{equation}
\partial_t \rho=i[\rho, \mathcal{H}]+\mathcal{L}[\rho],
\end{equation}
where the total Hamiltonian is $\mathcal{H}=\mathcal{H}_0+\mathcal{H}_{\text{Hol}}+\mathcal{H}_\text{d-d}$. The dissipative, incoherent dynamics stemming from the coupling of the electronic and vibrational degrees of freedom to their baths in thermal equilibrium, is included in the Lindblad part as a superoperator (an operator acting on density operators). A standard, diagonal superoperator in Lindblad form~\cite{breuer2002theory,weiss1999quantum,scully1997quantum,gardiner2004quantum,vogel2006quantum,walls2012quantum} is defined as
\begin{equation}
\label{Lstandard}
\mathcal{L}_\gamma[\rho] = \frac{\gamma_{\mathcal{O}}}{2}\left[ 2\mathcal{O} \rho(t)\mathcal{O}^{\dagger}- \mathcal{O}^{\dagger}\mathcal{O} \rho(t) - \rho(t)\mathcal{O}^{\dagger}\mathcal{O}\right],
\end{equation}
and describes decay at generic rate $\gamma_\mathcal{O}$ through a single channel with a generic collapse operator $\mathcal{O}$. The radiative dynamics stemming from the coupling of electronic transitions to the electromagnetic vacuum is, however, not in diagonal Lindblad form~\cite{lehmberg1970radiation} but achieves the following expression
\begin{equation}
\label{Lcoll}
\mathcal{L}_e[\rho] = \sum_{j,j'}\frac{\Gamma_{jj'}}{2}\left[ 2\sigma_{j} \rho\sigma^{\dagger}_{j'}- \sigma^{\dagger}_j\sigma_{j'} \rho - \rho\sigma^{\dagger}_j\sigma_{j'}\right].
\end{equation}
A simple diagonalization of the matrix of decay rates suffices to bring the expression above into standard Lindblad form and to see the emergence of $\mathcal{N}$ collective dissipation channels. The second contribution to $\mathcal{L}[\rho]$ stems from the non-radiative loss of vibrational excitation and is in standard Lindblad form with rate $\Gamma_\nu$ for all molecules and collapse operators $b_j$. This is an approximated model, as some care has to be taken regarding the correct collapse operator since the vibronic coupling can be strong ($\lambda\sim 1$) and the vibrational relaxation is typically much faster than the spontaneous emission. It can be shown that in such a case, a local master equation is thermodynamically not consistent and a global master equation approach has to be taken with correct collapse operator given by $b_j-\lambda\sigma^{\dagger}_j\sigma_j$ \cite{carmichael1973master, hu2015quantum, naseem2018thermodynamic}.
\section{Radiative properties of vibronically coupled emitters}
\label{Sec3}
The non-standard form of the radiative dissipation leads to cooperative effects which show the imprint of superradiance and subradiance. For ideal quantum emitters, such effects are well understood~\cite{lehmberg1970radiation} and analytically tackled e.g.~in Ref.~\cite{reitz2022cooperative}. However, the vibronic coupling appearing in the Hamiltonian in Eq.~\eqref{holstein} changes these effects considerably. We will focus on two distinct situations: i) inverted ensembles, where the dynamics is followed on the whole Bloch sphere and ii) the single excitation manifold, relevant under weak excitation conditions. We will make use of both an individual site basis (described by operators $\sigma_j$), as well as a collective basis, where symmetric and antisymmetric combinations of the $\sigma_j$ operators will be defined. We first show that vibrations lead to a degradation of the superradiant pulse emission in the Dicke limit. Then we analyze the symmetric/antisymmetric dynamics to show that both dissipative dynamics and vibronic effects lead to couplings among collective states of different symmetries. In the single excitation subspace, states of different symmetry do not couple via dissipative effects, allowing the derivation of a band structure describing the dispersion of excitations tunneling between molecules via the near field dipole-dipole interactions; this behavior is only changed owing to vibronic effects. \\
\subsection{Dissipation under vibronic coupling}
\indent Let us first review a few details on the vibronic coupling following the description in Ref.~\cite{reitz2019langevin}. For a single molecule indexed by $j$, the Holstein Hamiltonian can be diagonalized via a level-dependent polaron transformation $\mathcal{U}_j^\dagger=\ket{g}_j\bra{g}_j+\mathcal{D}_j^\dagger\ket{e}_j\bra{e}_j$ with the standard displacement operator defined as
\begin{equation}
\mathcal{D}_j=e^{-i\sqrt{2}\lambda p_j}=e^{\lambda(b_j^\dagger-b_j)}.
\end{equation}
In the polaron-displaced basis, the Holstein Hamiltonian $h_0^{(j)}+h_{\text{Hol}}^{(j)}$ becomes diagonal
\begin{equation}
\tilde{h}^{(j)}=\mathcal{U}_j^\dagger (h_0^{(j)}+h_{\text{Hol}}^{(j)})\mathcal{U}_j=\omega_0{\sigma}_j^\dagger{\sigma}_j+\nu {b}_j^\dagger{b}_j
\end{equation}
and has simple eigenvectors $\ket{g;n}_j$ and $\ket{e;n}_j$. The eigenvectors in the bare, original basis can be found by inverting the polaron transformation $\ket{g;n}_j$ and $\mathcal{D}\ket{e;n}_j$. The important property we have used is the transformation of the Pauli matrices under the vibrational displacement $\mathcal{U}_j^\dagger \sigma_j\mathcal{U}_j=\sigma_j \mathcal{D}_j$. The dressed operators describe polarons, i.e. hybrid electronic-vibrational operators. Furthermore we assume a thermal state with the average occupancy $\bar{n}=[\exp (\hbar \nu/(k_B T))-1]^{-1}$ (where $k_B$ is the Boltzmann constant). The partial trace over the vibronic displacement operators at temperature $T$ is therefore given by
\begin{equation}
\langle \mathcal{D}_j \mathcal{D}_{j'}^\dagger \rangle_T = e^{-\lambda^2(1+2\bar{n})(1-\delta_{j{j'}})}.
\label{thermal}
 \end{equation}
Note that at $T=0$ the above trace reduces simply to $\langle \mathcal{D}_j \mathcal{D}_{j'}^\dagger \rangle_{T=0} = \mathrm{exp}[-\lambda^2(1-\delta_{j{j'}})]$ giving unity on a given molecule but a reduction by the Franck-Condon factor $e^{-\lambda^2}$ for a two molecule term.\\
\indent We can now apply the polaron transformation with an operator $\mathcal{U}^\dagger=\textstyle \prod_j \mathcal{U}_j^\dagger$ such as to diagonalize the whole vibronic Hamiltonian. We are however left with the polaron transformed dipole-dipole interaction
as well as a polaron transformed Lindblad term, which describes dissipation via polaron collapse operators
\begin{align}
\label{Lcollpolaron}
\tilde{\mathcal{L}}_e[\rho] = \sum_{j,j'}\frac{\Gamma_{jj'}}{2}\left[ 2\sigma_{j} \mathcal{D}_j\rho \mathcal{D}_{j'}^\dagger \sigma^{\dagger}_{j'}-\{
\mathcal{D}_j^\dagger\sigma^{\dagger}_j\sigma_{j'}\mathcal{D}_{j'},\rho   \}\right],
\end{align}
where the last term denotes an anticommutator. We will then make the assumption that the vibrations are in a thermal state and that the electronic and vibrational states factorize. This leads to a renormalization of the dipole-dipole interaction $\Omega_{jj'}^\lambda=\Omega_{jj'}e^{-\lambda^2(1+2\bar{n})}$ as well as renormalized off-diagonal (or mutual) decay rates  as evident from the polaron transformed Lindblad term
\begin{align}
\label{Lindbladtraced}
\tilde{\mathcal{L}}_e[{\rho}] &= \sum_{jj'} e^{-\lambda^2 (1+2\bar{n})(1-\delta_{jj'})} \frac{\Gamma_{jj'}}{2}\Big[2 {\sigma}_j \rho {\sigma}^\dagger_{j'} - \{ {\sigma}^\dagger_j {\sigma}_{j'}, \rho \} \Big].
\end{align}
Notice that for large $\lambda$ or large thermal occupancies, the off-diagonal elements of the Lindblad term above (corresponding to cooperative emission) vanish, leading to the disappearance of any subradiant or superradiant behavior and the recovery of the independent decay behavior.

\subsection{Dynamics on the Bloch sphere}
\label{bloch}
\begin{figure}[t]
\includegraphics[width=0.99\columnwidth]{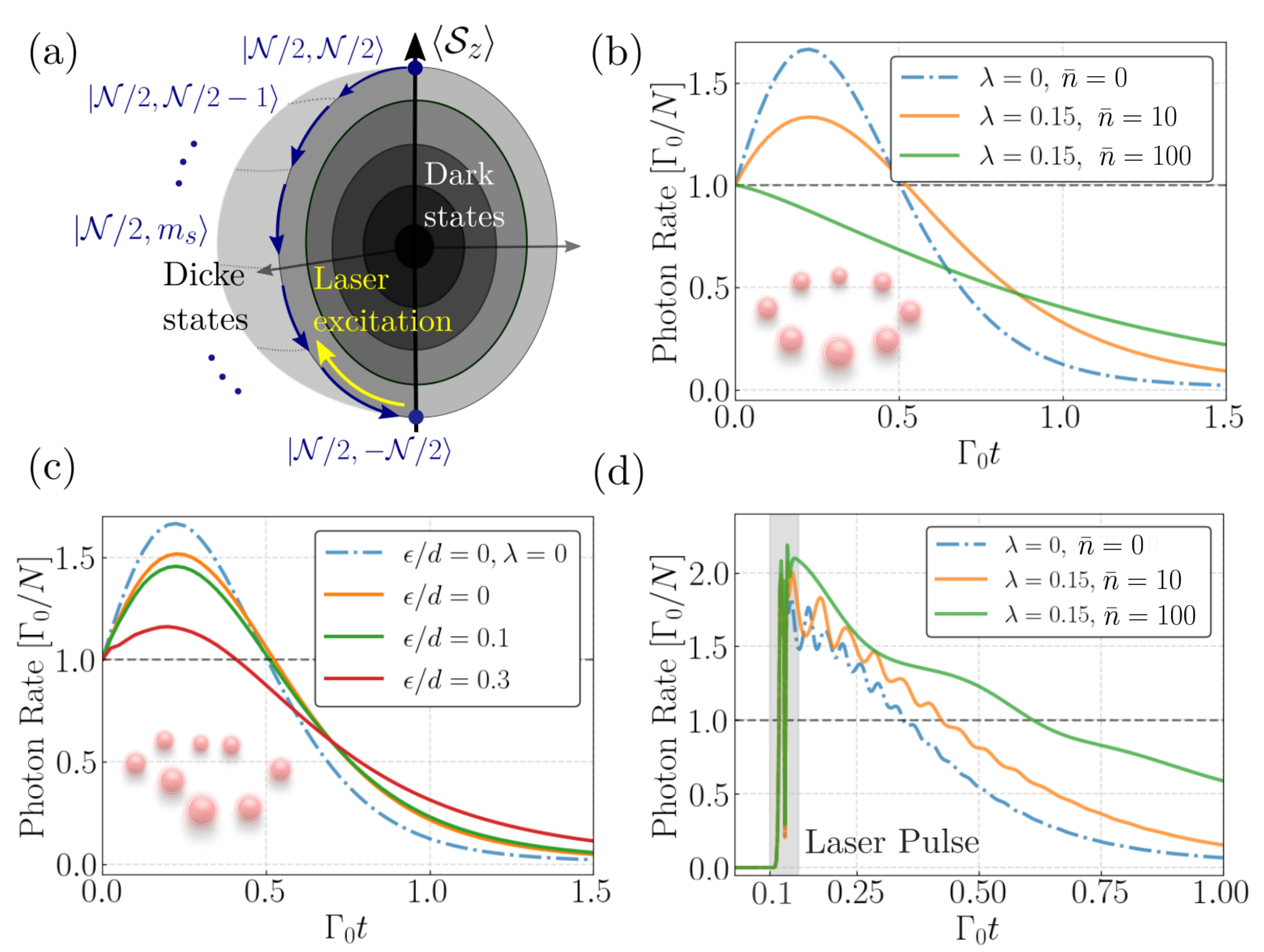}
\caption {(a) Illustration of the collective Bloch sphere for $\mathcal{N}$ emitters. The symmetric subspace is spanned by $\mathcal{N}+1$ Dicke states, while the inside of the sphere is spanned by antisymmetric states. (b) Superradiant decay for an initially fully inverted ring of $\mathcal{N} = 8$ molecules with their vibrational degree of freedom in thermal equilibrium at various temperatures. (c) Scaling of the superradiant pulse intensity for a fully inverted system of molecules in the ring configuration, as a function of increasing positional disorder $\epsilon$ and with vibronic coupling $\lambda=0.15$ (d) Time dependence of the intensity of emission for a ring of $\mathcal{N} = 8$ molecules driven by a laser pulse with the frequency matched to the symmetric state resonance $\omega_\ell = \omega_\mathcal{S}$. The inter-molecular separation in all plots is $d=0.04\uplambda_0$ and the dipoles are linearly polarized perpendicular to the plane of the ring. Parameters are fixed to $\eta = 260 \Gamma_0, t_0 = 0.1 / \Gamma_0$ and $\tau = 0.1/\Gamma_0$.}
\label{fig2}
\end{figure}

We will first analyze the standard Dicke superradiance phenomenon extended to the case of molecules, i.e. for a vibronic coupling characterized by a non-zero Huang-Rhys factor $\lambda=0$. To this end, we will make use of a Bloch sphere representation for the collective spin of the system as illustrated in Fig.~\ref{fig2}(a). We use of a non-standard angular momentum representation for the sum of $\mathcal{N}$ spin $1/2$ subsystems where a collective collapse operator is introduced as a symmetric combination $\mathcal{S}=\sum_j \sigma_j/\sqrt{\mathcal{N}}$. The Cartesian components are $\mathcal{S}_z=\sum_j \sigma_j^{(z)}$, $\mathcal{S}_x=\mathcal{S}+\mathcal{S}^\dagger$ and  $\mathcal{S}_y=-i(\mathcal{S}-\mathcal{S}^\dagger)$. Common eigenstates of the total spin vector $\vec{\mathcal{S}}$ and $\mathcal{S}_z$ are then denoted by $\ket{s,m}$ where the quantum number $s$ runs $0$ or $1/2$ to $\mathcal{N}/2$ and $m$ from $-s$ to $s$. In the symmetric subspace the so-called Dicke states arise denoted by $|\mathcal{N}/2,m\rangle$ and obtained by fixing $s$ to its maximal value $\mathcal{N}/2$. The action of the lowering/raising operators on the Dicke states is $\mathcal{S}|\mathcal{N}/2,m\rangle=\alpha_m^{(-)}|\mathcal{N}/2,m-1\rangle$ and $\mathcal{S}^\dagger|\mathcal{N}/2,m\rangle=\alpha_m^{(+)}|\mathcal{N}/2,m+1\rangle$ where the coefficients are
\begin{align}
\alpha_m^{(\pm)} = \frac{1}{\sqrt{\mathcal{N}}}\sqrt{(\mathcal{N}/2\mp m)(\mathcal{N}/2\pm m+1)}.
\end{align}
In the Dicke limit ($d=0$) and in the absence of vibrations ($\lambda=0$), the Lindblad term in Eq.~\eqref{Lindbladtraced} can be immediately diagonalized as a single loss channel with collapse operator $\mathcal{S}$ at superradiant rate $\mathcal{N}\Gamma_0$. This is no longer when $\lambda \neq 0$ or $d>0$ or both, as population spills outside the symmetric subspace towards the interior of the Bloch sphere. This behavior can be easily understood in a collective basis, where additional $\mathcal{N}-1$ antisymmetric operators are introduced
\begin{align}
\mathcal{A}_k &=\frac{1}{\sqrt{\mathcal{N}}}\sum_{j=1}^{\mathcal{N}}\sigma_j e^{2 \pi i j k /\mathcal{N}}, \quad \text{for} \quad k \in \{1,...,\mathcal{N}-1 \},
\end{align}
under the requirement that they are orthogonal to the $\mathcal{S}$ operator (this can be done more generally, for example, via a Gram-Schmidt algorithm).
The Hamiltonian can be easily diagonalized in terms of collective operators giving
\begin{equation}
\mathcal{H} = \omega_\mathcal{S}^\lambda \mathcal{S}^\dagger \mathcal{S} + \sum_{k=1}^{\mathcal{N}-1} \omega_k^\lambda \mathcal{A}_k^\dagger \mathcal{A}_k.
\end{equation}
This is based on the orthonormality condition $\sum_{k=1}^\mathcal{N}e^{2\pi i k (j-j')/\mathcal{N}}=\mathcal{N}\delta_{jj'}$ and on the cyclic symmetry of the ring allowing to write any double sums $\sum_{j'\neq j}e^{2\pi i k (j-j')/\mathcal{N}} \Omega_{jj'}=\sum_{j'=2}^\mathcal{N}e^{2\pi i k (j'-1)/\mathcal{N}} \Omega_{1j'}$. The modified eigenenergies are given by $\omega_\mathcal{S}^\lambda=\omega_0+\sum_{j=2}^{\mathcal{N}} \Omega_{1j}^\lambda$, for the symmetric states and
\begin{equation}
\omega_k^\lambda=\omega_0+\sum_{j=2}^{\mathcal{N}} \Omega_{1j}^\lambda e^{2\pi i(j-1)k/\mathcal{N}},
 \end{equation}
for the set of antisymmetric combinations.\\
\indent The thermally averaged Lindblad term from Eq.~\eqref{Lindbladtraced} can now be split into $\mathcal{N}$ decay channels: a symmetric one with loss rate $\Gamma_\mathcal{S}^\lambda(d)$ and $\mathcal{N}-1$ antisymmetric channels with rates $\Gamma_k^\lambda(d)$. These can be expressed as
\begin{align}
\Gamma^{\lambda}_{\mathcal{S},k}(d)=\Gamma_0\left[1-e^{-\lambda^2 (1+2\bar{n})}\right]+\Gamma^{\lambda= 0}_{\mathcal{S},k}(d)e^{-\lambda^2 (1+2\bar{n})},
\end{align}
in terms of the bare rates for zero vibronic coupling $\Gamma^{\lambda= 0}_\mathcal{S}(d)=\textstyle \sum_{j=1}^{\mathcal{N}}\Gamma_{1j}(d)$ and $\Gamma^{\lambda= 0}_k(d)=\textstyle \sum_{j=1}^{\mathcal{N}}\Gamma_{1j}(d) e^{i 2\pi (j-1) k /\mathcal{N}}$. Notice that for zero distance and no vibronic couplings, we recover the Dicke superradiance effect with rate $\mathcal{N}\Gamma_0$. For larger distances this is effect is reduced; additional reduction appears for nonzero vibronic coupling and temperature. Finally, for large $\lambda$ or $\bar{n}$, a complete washout of superradiance occurs and the first term in the expression above indicates the independent rate $\Gamma_0$ for both symmetric and antisymmetric states.\\
\begin{figure*}[t]
\includegraphics[width=0.9\textwidth]{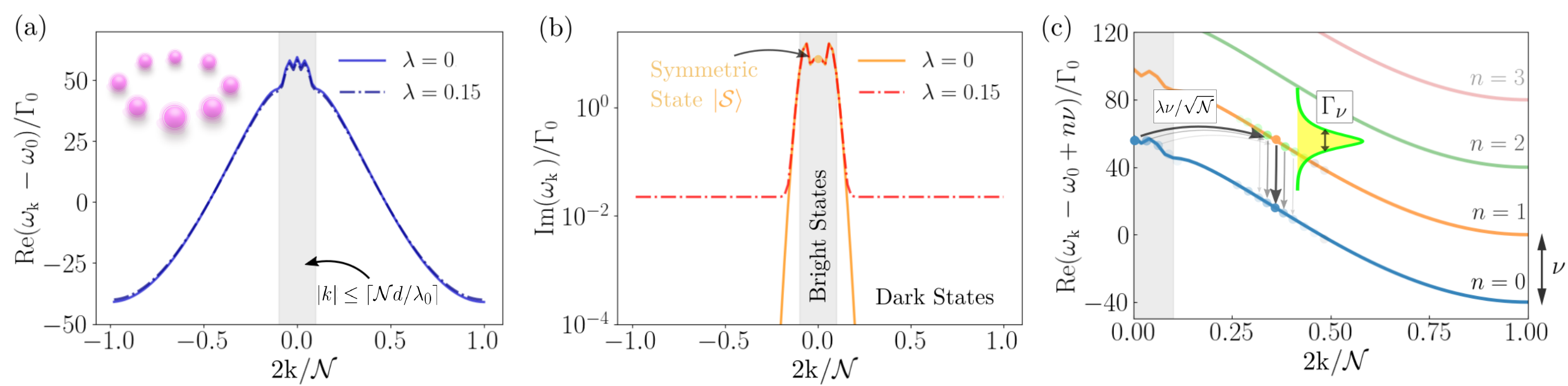}
\caption {(a) Single-excitation dispersion relation in the first Brillouin zone for a ring of $\mathcal{N}=100$ transversely polarized molecules at zero temperature with nearest-neighbour separation $d=0.05\uplambda_0$. Bright states enclosed by the shaded region are characterized by a mode number $|k|\le \lceil \mathcal{N}d/\uplambda_0\rceil$ whereas the region beyond is occupied by dark states. (b) Bright states feature a finite decay rate with the symmetric state located at $k = 0$. For $d\ll \uplambda_0$ the dark state decay rates are approximately given by $\Gamma_{k}^\lambda\sim (1-e^{-\lambda^2})\Gamma_0$ whereas the bright state decay rate approaches $\Gamma_{\mathcal{S}}^\lambda\sim \Gamma_0 + e^{-\lambda^2}(\mathcal{N}-1)\Gamma_0$ . (c) Dispersion curves in the full Hilbert space (for $\lambda=0.15$). States with $n$ vibrational energy quanta are shifted by $n \nu$ with respect to the zero-vibrational states. Vibrations lead to coherent population transfer from bright states with lower vibrational quantum state excitation to dark states with higher vibrational quantum state excitation, at a coupling strength $\lambda \nu /\sqrt{\mathcal{N}}$. The process is followed by non-radiative vibrational relaxation into the dark state with zero vibrations at a rate $\Gamma_\nu\gg \Gamma_0$.}
\label{fig3}
\end{figure*}
\indent For a better understanding of the coupling between states of different symmetries, we now perform an analysis in the full Hilbert space, i.e.~without tracing over the thermal bath. Instead, intuitive understanding is offered by an additional transformation to a collective basis for the vibrational degrees of freedom as well, introduced via
\begin{align}
Q_k = \frac{1}{\sqrt{\mathcal{N}}} \sum_{j=1}^\mathcal{N} e^{2 \pi i j k /\mathcal{N}} (b_j + b_j^\dagger),
\end{align}
where $k \in \{1,...,\mathcal{N} \}$ (with $k=\mathcal{N}$ corresponding to the symmetric vibrational mode) and with the momentum quadratures satisfying $[Q_k,P_{k'}]=2i\delta_{k{k'}}$. An interaction term emerges, coupling the symmetric state to the antisymmetric manifold
\begin{align}
  \label{symmetric-coupling}
\mathcal{H}_{\text{int}}^{\mathcal{S}\mathcal{A}}=- \frac{\lambda\nu}{\sqrt{\mathcal{N}}} \sum_{k=1}^{\mathcal{N}-1} (Q_k \mathcal{S}^\dagger \mathcal{A}_k + h.c.),
\end{align}
via the position quadratures of the collective vibrations. This coupling is responsible for the spilling of population into the interior of the Bloch sphere even when fully symmetric driving for the system takes place. The effect will be useful in order to understand the dynamics of the coherent nanoscale source analyzed in Sec.~\ref{Sec5}. In addition, couplings within the antisymmetric states manifold emerge via
\begin{align}
  \label{antisymmetric-coupling}
\mathcal{H}_{\text{int}}^{\mathcal{A}\mathcal{A}}=-\frac{\lambda\nu}{\sqrt{\mathcal{N}}} \sum_{k\neq k'}^{\mathcal{N}-1}(Q_{k-k'}\mathcal{A}_k^\dagger \mathcal{A}_{k'} +h.c.).
\end{align}
This Hamiltonian shows a redistribution of energy within the whole manifold of antisymmetric states. In the mesoscopic limit, a very large number of such states exist, leading to a quick energy loss from the symmetric subspace to all other subspaces orthogonal to it. This observation could constitute the basis for an effective theory as developed in Ref.~\cite{sommer2020molecular}, which allows for the derivation of an effective unidirectional Markovian loss dynamics for the symmetric operator.\\
\indent Let us now numerically illustrate the Dicke superradiant behavior for a tightly packed system of emitters and check the analytically obtained results. We depart now from the ideal case of zero separation, and consider a ring of $\mathcal{N}=8$ molecules with separation of $d=0.04\uplambda_0$. The inclusion of the inherent coherent dipole-dipole interactions leads to a shift of the collective symmetric state which we effectively target in the numerical simulations. These results are illustrated in Fig.~\ref{fig2}(b) as a function of the environmental temperature for $\lambda=0.15$. One can clearly observe the washing out of the standard Dicke superradiant pulsed decay, plotted as the intensity of the emitted pulse as a function of time. In the large temperature limit, the independent decay behavior is recovered, signaling that temperature effects hinder the build up of two-particle correlations necessary for the emergence of superradiant behavior. In Fig.~\ref{fig2}(c), some robustness to positional disorder is observed where each molecule is randomly displaced around its equilibrium position by a normal distribution of standard deviation $\epsilon$. The trajectories are plotted after performing an average over 100 disorder realizations with $\lambda = 0.15$.\\
\indent Finally, we numerically illustrate time dynamics under resonant laser drive ($\omega_\ell=\omega_\mathcal{S}$), modeled by a pulsed excitation with electric field amplitude
\begin{align}
E_\mathrm{in}(t) = \Omega_\ell(t)\sum_{j=1}^\mathcal{N}\Big(e^{-i \vec{k}_\ell \cdot \vec{r}_j} e^{i\omega_\ell t}\sigma_j + e^{i \vec{k}_\ell \cdot \vec{r}_j} e^{-i\omega_\ell t}\sigma_j^\dagger\Big).
\label{gauss}
\end{align}
The laser pulse is considered to be impinging from the $xy$-plane with a linear polarization $\hat{e}_z$ coinciding with the dipole orientation of the molecules. The time dependence is a Gaussian envelope of the form $\Omega_\ell(t) = \eta \ \mathrm{exp}[-(t-t_0)^2/\tau^2]$, with maximum amplitude $\eta$ and duration $\tau$ and the wave vector of the laser is assumed to be $\vec{k}_\ell = k_0 \hat{e}_x$. The situation is depicted in Fig.~\ref{fig2}(d) and shows that superradiant emission is reached even at large temperatures, via the properly tailored pulsed, resonant addressing. In fact, vibronic coupling not only leads to an increase of possible states reachable by the laser which in the Dicke regime ($d \ll \uplambda_0$) would otherwise be prohibited, but additionally decreases the dephasing stemming from the coherent dipole-dipole interaction and thereby leads to an increased photon emission after the pulse is switched off.
\indent Let us now use our approach to compare our results to analytical predictions~\cite{masson2021universality,robicheaux2022} which show that superradiant decay of a fully inverted ensemble of two-level emitters can be predicted purely by the geometry of the system by observing that a positive slope of the total emitted intensity $\sum_{ij}\Gamma_{ij}\langle \sigma^\dagger_i \sigma_j \rangle(t)$ at $t=0$ is a good criterion for superradiant emission. The condition derived in these references reads $\sum_{k=1}^{{\mathcal{N}}} \Gamma_k^2 > 2  \mathcal{N} \Gamma_0^2$, where $\Gamma_k$ are the collective decay rates corresponding in our case to a fixed distance $d$ and zero vibronic coupling. This can be immediately translated to the case of $\mathcal{N}$ identical molecules, where the factor $\mathrm{exp}[{-\lambda^2}(1+2\bar{n})]$ is crucial, leading to the following condition for the emergence of superradiant decay
\begin{align}
 \sum_{k=1}^{{\mathcal{N}}} \left(\Gamma_k^{\lambda=0}\right)^2 > \frac{1+e^{-2\lambda^2(1+2\bar{n})}}{e^{-2\lambda^2(1+2\bar{n})}}\mathcal{N} \Gamma_0^2.
\end{align}
This shows that with increasing temperature and/or vibronic coupling the condition for superradiance to occur is more difficult to meet and in the Dicke limit only one collective decay rate is non-zero and the inequality reduces to
\begin{equation}
 \mathcal{N} >  \frac{1+e^{-2\lambda^2(1+2\bar{n})}}{e^{-2\lambda^2(1+2\bar{n})}},
\end{equation}
which sets an upper bound of $\lambda^2(1+2\bar{n}) < \mathrm{log}(\mathcal{N}-1)/2$ for the Huang-Rhys factor $\lambda^2$. It then follows that for molecular systems at zero temperature and in the Dicke limit ($d/\uplambda_0 = 0$), the criteria for the Huang-Rhys factor for which superradiant effects can still be observed is $\lambda^2 < \log(\mathcal{N}-1)/2$.

\subsection{Dynamics in the single excitation subspace}
\label{singleex}

The single excitation subspace is especially relevant for the case of mesoscopic systems of quantum emitters driven with a very weak excitation pulse. Dipole-dipole interactions induce tunneling behavior between neighboring emitters, allowing the understanding of the system's properties in terms of the band structure or dispersion relations for the propagation of collective excitations. Non-hermitian, dissipative effects such as superradiance and subradiance of such linear systems can also be understood in terms of the localization of collective states within or outside a light cone.\\
\indent Restricting the Hilbert space to a single excitation, one can recast the Hamiltonian in Eq.~(\ref{Dipole-Ham}) into the following non-Hermitian form (by disregarding the recycling term in the Lindbladian)
\begin{align}
   \mathcal{H}= \omega_0 \sum_{j}^\mathcal{N} \sigma_j^\dagger \sigma_j + \sum_{jj'}^\mathcal{N} \Big[\Omega^\lambda_{jj'}(d)-i\frac{\Gamma^\lambda_{jj'}(d)}{2} \Big] \sigma_j^\dagger \sigma_{j'}.
\end{align}
As mentioned in the previous subsection, the collective basis offers a diagonalization of the dynamics. As opposed to the full Bloch sphere case, in the single excitation one can proceed with diagonalization of both coherent and incoherent parts by writing $\mathcal{H} = \sum_{k=1}^{\mathcal{N}} \bar{\omega}_k^\lambda \mathcal{A}^\dagger_k \mathcal{A}_k$ where by definition the symmetric state operator corresponds to the case $\mathcal{S}=\mathcal{A}_{k=\mathcal{N}}$. The eigenenergies and the decay rates are given by the real and imaginary part of the complex eigenvalues
\begin{align}
\bar{\omega}_{k}^\lambda = \omega_0 + \Omega^\lambda_{k}(d)-i \frac{\Gamma^\lambda_{k}(d)}{2}.
\end{align}
The excitations can be understood in terms of the quasimomentum $q=2\pi k/(\mathcal{N}d)$, where due to the periodicity we can define the first Brillouin zone by the index $k = 0,\pm 1,... \pm \lceil (N-1)/2 \rceil$ where $ \lceil x\rceil$ denotes the ceiling function. Note that the center of the Brillouin zone $k=0$ corresponds to the symmetric mode and the edges at $k_\pm=\pm \lceil ( \mathcal{N}-1)/2 \rceil$ to the most subradiant modes with degenerate eigenvalues $\omega_{k_\pm}$.
This can be understood from the wave equation $q^2 + q_\perp^2  = (2\pi/\uplambda_0)^2$ which requires that for modes with $|q|\ge 2\pi/\uplambda_0$, the radial electric field components are evanescent, i.e., exponentially decaying, and the excitation is guided along the ring. Modes inside the region $|q|\le 2\pi/\uplambda_0$ on the other hand have electric field components transverse to the ring and are therefore radiating energy away into the vacuum. \\
\indent In Fig.~\ref{fig3}(a) the dispersion relation for a ring of $\mathcal{N} = 100$ molecules is shown for the cases with and without vibronic coupling $\lambda$. The region defined by the integer number $|k|\ge \lceil \mathcal{N}d/\uplambda_0\rceil$ is occupied by dark states as shown in Fig.~\ref{fig3}(b) whose decay rates (for a fixed $k$) are decreasing exponentially with the number of emitters for $\lambda=0$ \cite{asenjo2017exponential}. For molecules with a non-zero vibronic coupling $\lambda$ the exponential scaling gets strongly modified and the decay rates for eigenstates with mode number $k$ are approximately given by $\Gamma_k^\lambda/\Gamma_0 \sim (1-e^{-\lambda^2})$.\\
\indent While Figs.~\ref{fig3}a and b show the real and imaginary parts of the dispersion relation using the reduced Hamiltonian, one can also discuss the dispersion relation in the collective basis including vibrations. The term in Eq.~\eqref{symmetric-coupling} illuminates the fact that the presence of vibrations causes a coherent transfer of population between the symmetric mode $\mathcal{S}$ and the dark modes $\mathcal{A}_k$. The coupling strength between the symmetric mode in the vibrational ground state and a dark mode with one vibrational excitation is given by $\lambda \nu / \sqrt{\mathcal{N}}$ which is illustrated in Fig.~\ref{fig3}(c). Since the vibrational relaxation rate $\Gamma_\nu $ is fast compared to the timescale $1/\Gamma_0$ of the electronic decay rate, the population relaxes quickly to the dark state with no vibrational quanta. In general the coupling strength between $\mathcal{S}$ with $n$ vibrations and mode $\mathcal{A}_{k}$ with $n+1$ vibrations is given by $\bra{\mathcal{S},n} \mathcal{H}\ket{\mathcal{A}_{k},n+1} = \sqrt{(n+1)/\mathcal{N}}\lambda \nu$ where the Hamiltonian includes the vibrational degrees of freedom and in particular the terms in Eq.~\eqref{symmetric-coupling}-\eqref{antisymmetric-coupling} which mediate the coherent transfer.
The rate of transfer for a ring geometry is derived in section \ref{Sec4} and generally the large vibrational linewidth $\Gamma_\nu \gg \Gamma_0$ will create resonances between multiple modes thereby enhancing the population transfer to the dark state manifold.
\begin{figure}[b]
\includegraphics[width=0.92\columnwidth]{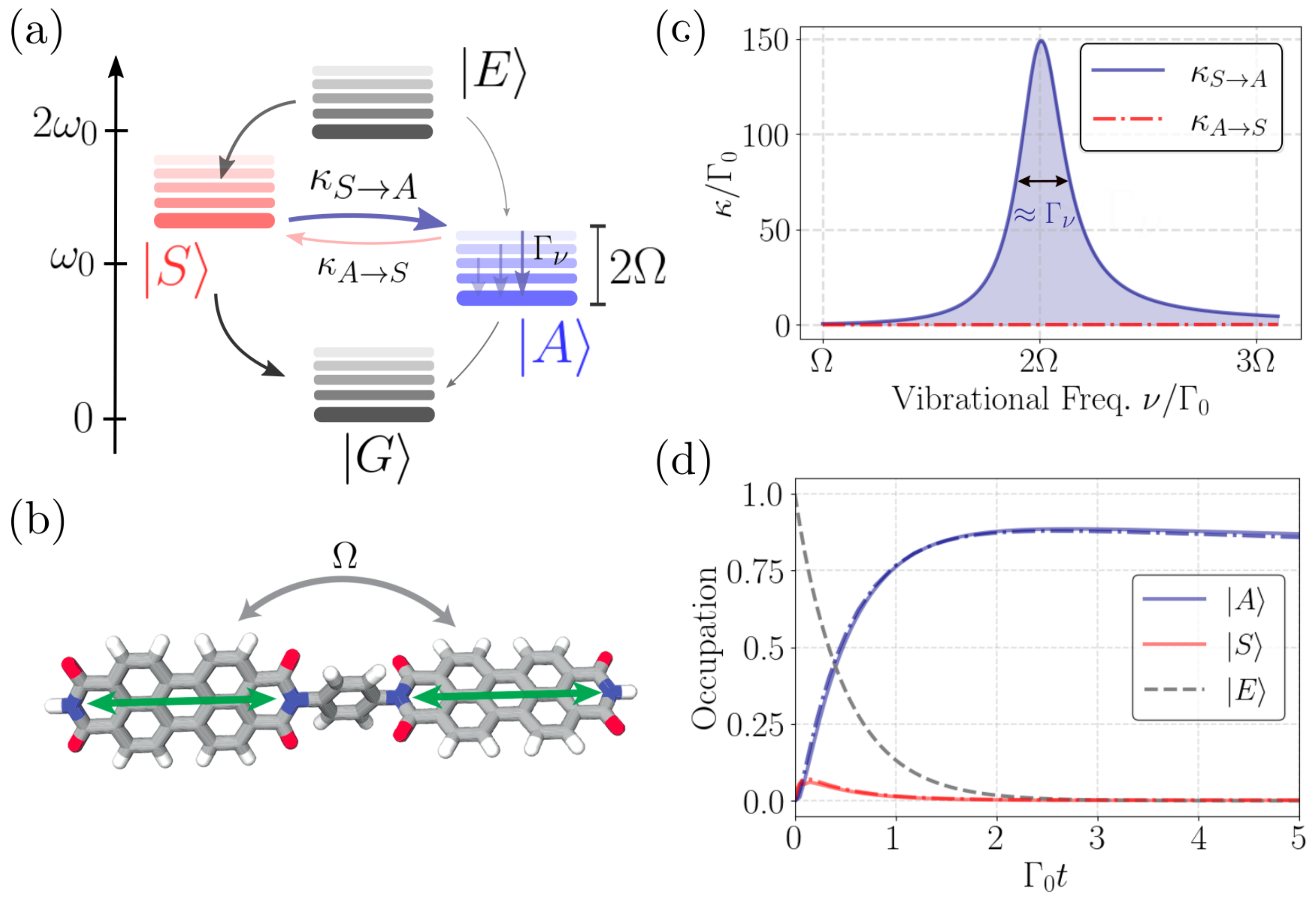}
\caption {(a) Energy diagram showing population transfer between symmetric (superradiant) and antisymmetric (subradiant) collective states via their mutual coupling to the vibrational bath. (b) Illustration of a molecular dimer where two identical chromophores are separated by an insulating bridge. Energy transfer between the two chromophores can take place via near field coupling on length on the order of nanometers. The situation depicted here shows in-plane dipoles (resulting in $\Omega<0$). (c) Energy transfer rates between the symmetric and antisymmetric dimer state as a function of the vibrational frequency. (d) Time evolution of a fully inverted molecular dimer. The fully excited state decays exponentially via the symmetric state which transfers energy to the antisymmetric state. The analytical results in dashed-dotted lines show a good agreement. Parameters are $\nu = 2\Omega, d = \uplambda_0/40 , \Gamma_\nu = 30\Gamma_0,\lambda = 0.1, \Omega(d) \approx 191.1 \Gamma_0$ and polarization perpendicular to dimer axis.}
\label{fig4}
\end{figure}

\section{Subradiant state preparation in molecular dimers and rings}
\label{Sec4}

Molecular dimers are ideal for the study of dipole-dipole induced energy shifts at very small separations and for the study of the interplay between electronic and vibrational quantum superpositions~\cite{ halpin2014two}. In such compounds, two chromophores are linked by insulating bridges which do not allow for charge migration and do not shift the bare electronic transitions. In Ref.~\cite{diehl2014emergence}, an experimental study of molecular dimers shows the possibility to control the inter-chromophoric distance from $1.3\,\mathrm{nm}$  to $2.6\,\mathrm{nm}$  while keeping the orientation of each chromophore dipole fixed.  Previous theoretical studies have focused mainly on the purely coherent interactions and have neglected the effects of vibrational relaxation and collective spontaneous emission \cite{merrifield1963vibronic, eisfeld2005vibronic, chenu2013enhancement}.\\
\indent Here, we show that the coupling between symmetric (bright) and antisymmetric (dark) collective states in a vibronic dimer, combined with the vibrational relaxation can lead to an efficient preparation of long-lived quantum entangled states of the two chromophores (see Fig.~\ref{fig4}). The mechanism is reminiscent of the process of FRET (F{\"o}rster resonance energy transfer) between donor and acceptor molecules, where coherent energy exchanges followed by quick vibrational relaxation can lead to a unidirectional flow of energy.\\
\indent The model is described by the free Hamiltonians $h_0^{(1)}+h_0^{(2)}$ to which we add $h_{\text{Hol}}^{(2)}+h_{\text{Hol}}^{(2)}$ and the two-particle term
$\mathcal{H}_{\text{d-d}}=\Omega(\sigma_1^\dagger\sigma_2+\sigma_2^\dagger\sigma_1)$ describing  excitation exchange between the two chromophores via the near field dipole-dipole coupling. We make use of the collective basis representation with $\mathcal{S}=(\sigma_1+\sigma_2)/\sqrt{2}$ as the symmetric operator and a single antisymmetric, orthogonal operator $\mathcal{A}=(\sigma_1-\sigma_2)/\sqrt{2}$. We define collective vibrational quadratures $Q_{\pm}=(q_1\pm q_2)/\sqrt{2}$ and $P_\pm=(p_1\pm p_2)/\sqrt{2}$ as well. The free Hamiltonian of electronic and vibrational degrees of freedom then can be expressed as
\begin{align}
\label{dimer}
\mathcal{H}_0^\text{dim}=\omega_\mathcal{S} \mathcal{S}^\dagger \mathcal{S} +\omega_\mathcal{A} \mathcal{A}^\dagger \mathcal{A}+\frac{\nu}{4}\sum_{k=\pm}(Q_k^2+P_k^2),
\end{align}
where the collective states frequencies $\omega_\mathcal{S}=\tilde{\omega}_0+\Omega-\lambda\nu Q_+/\sqrt{2} $ and $\omega_\mathcal{A}=\tilde{\omega}_0-\Omega-\lambda\nu Q_+/\sqrt{2} $ become now operators which include the symmetric vibrational coordinate. The energy scheme of the dimer is presented in Fig.~\ref{fig4}(a) showing vibrationally-dressed collective electronic states. While in the absence of motion the symmetric and antisymmetric states are orthogonal to each other, this is no longer the case when vibrations are included allowing for transitions between them. The vibrational degrees of freedom then couple the two states via the relative motion coordinate $Q_-$
\begin{align}
\label{intdimer}
\mathcal{H}_\text{int}^\text{dim}=-\frac{\lambda\nu}{\sqrt{2}} Q_- (\mathcal{S}^\dagger\mathcal{A}+\mathcal{A}^\dagger \mathcal{S}),
\end{align}
such that the total dimer Hamiltonian expresses as $\mathcal{H}_0^\text{dim}+\mathcal{H}_\text{int}^\text{dim}$.
The interaction term in Eq.~(\ref{intdimer}) can mediate transfer of excitation between the bright and dark state through the annihilation or creation of a vibrational quantum of the relative motion coordinate.
\begin{figure}[t]
\includegraphics[width=0.99\columnwidth]{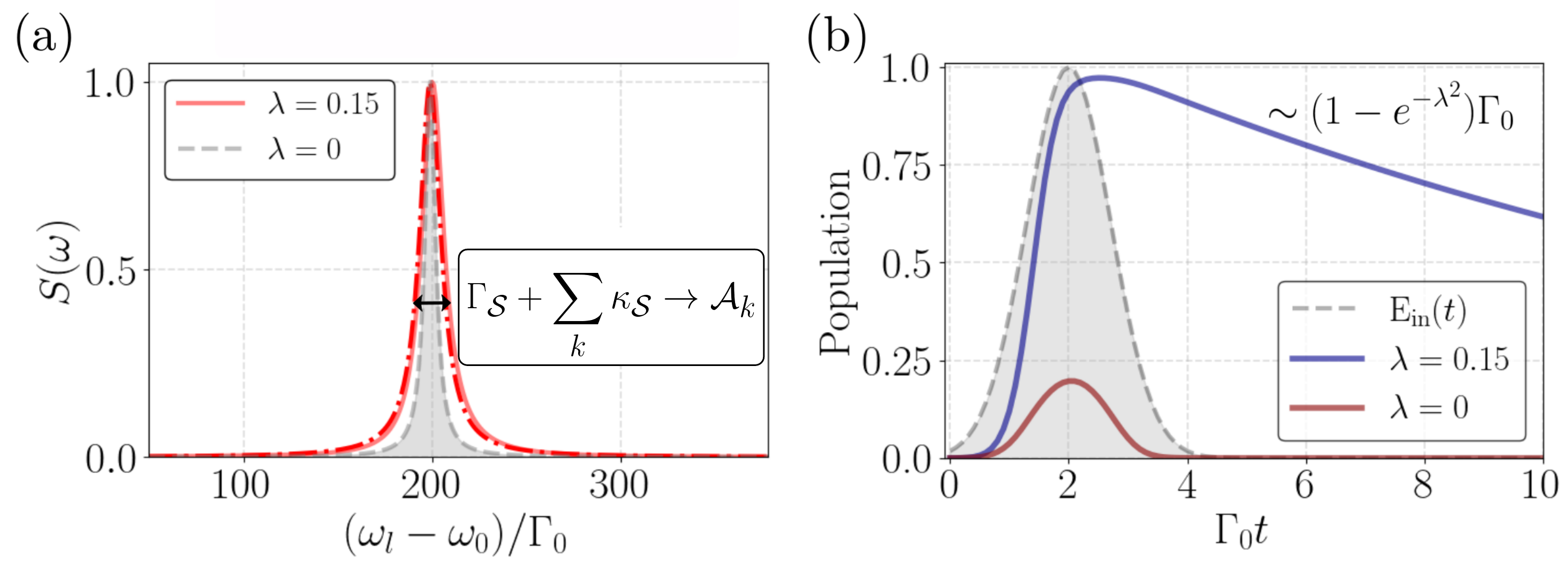}
\caption {(a) Absorption spectrum in steady state for a ring with $\mathcal{N}=7$ molecules with the linewidth of the symmetric state broadened by the sum of the energy transfer rates to the dark state manifold. The dashed-dotted line is a Lorentzian with linewidth given by $\Gamma_\mathcal{S}+\sum_k\kappa_{\mathcal{S}\to\mathcal{A}_k}$ and maximum at $\omega_\ell= \omega_0+\Omega_\mathcal{S}^{\lambda=0}$. (b) Laser pulse with a Gaussian time profile as in Eq.~(\ref{gauss}) with $\eta =2.5 \ \Gamma_0, t_0 = 2/\Gamma_0, \tau = 1/\Gamma_0$. The laser frequency $\omega_\ell$ is tuned to the superradiant mode $k=0$. The single excitation manifold is populated almost with unity and decays with a subradiant rate $\sim \Gamma_0(1-e^{-\lambda^2})$ afterwards. Further parameters for both plots are  $d = \uplambda_0/30 , \Gamma_\nu = 100\Gamma_0,\lambda = 0.15,\nu = 120 \Gamma_0$.}
\label{fig5}
\end{figure}
Under the assumption that the vibrational relaxation is fast as compared to the coherent coupling $\Gamma_\nu\gg\lambda\nu$ as well as all other decay rates, a perturbative set of rate equations for the populations $p_\calS=\braket{\calS^\dagger\calS}$ and $p_\calA=\braket{\calA^\dagger\calA}$ can be obtained (for derivation see Appendix \ref{C})
\begin{subequations}
\begin{align}
\dot{p}_\calS &=-(\Gamma_\calS+\kappa_{\calS\to\calA})p_\calS +\kappa_{\calA\to\calS}p_\calA  ,\\
\dot{p}_\calA &=-(\Gamma_\calA+\kappa_{\calA\to\calS})p_\calA +\kappa_{\calS\to\calA}p_\calS ,
\end{align}
\end{subequations}
with transfer rate from the symmetric to antisymmetric state
\begin{align}
\kappa_{\calS\to\calA}=\frac{\lambda^2\nu^2\Gamma_\nu/2}{(\Gamma_\nu/2)^2+(2\Omega-\nu)^2}.
\end{align}
The transfer from the antisymmetric to symmetric state $\kappa_{\calA\to\calS}$ has a similar expression, however with a term $(2\Omega+\nu)$ present in the denominator. For $\Omega>0$ the resonance condition is given by $2\Omega=\nu$ leading to unidirectional transfer from the symmetric to the antisymmetric state while the back transfer is off-resonant and therefore suppressed [see Fig.~\ref{fig4}(c)]. In Fig.~\ref{fig4}(d) we plot the time dynamics of a dimer initialized in the fully excited state $\ket{E}$ under this resonance condition. Initial decay to the symmetric state is followed immediately  by a rapid transfer to the antisymmetric state, causing only a small temporary population in the symmetric state and a large accumulation of population in the antisymmetric state. Remarkably, this can lead to a near-unity population in the antisymmetric state even for moderate vibronic coupling strengths $\lambda$. Since vibrational frequencies are on the order of $\nu/2\pi\sim10\,\mathrm{THz}$ and the spontaneous emission rate is on the order of $\Gamma_0/2\pi\sim10\,\mathrm{MHz}$, this resonance condition requires dipole-dipole shifts on the order of $\sim 10^6\,\Gamma_0$ which can be achieved by dimers with $\mathrm{nm}$ separations.\\
\indent Let us finally remark that the dark state preparation scheme described here for the dimer can be extended to configurations of many molecules in the ring configuration. To this end we have performed numerical simulations showing the drive of collective states which are not accessible via direct illumination but are populated via the incoherent, vibrationally mediated transfer. In Fig.~\ref{fig5}(a), the enhanced absorption profile for a ring of $\mathcal{N}=7$ molecules signals the transfer of population from the symmetric, laser accessible collective state to a number of initially dark states. The increase in the linewidth is simply given by the sum of all transfer rates to the dark state manifold which are obtained as a generalization of the dimer result
\begin{align}
\sum_k\kappa_{\mathcal{S}\to\mathcal{A}_k}=\sum_k\frac{\lambda^2\nu^2\Gamma_\nu/2}{(\Gamma_\nu/2)^2+(\Omega_\mathcal{S}-\Omega_{\mathcal{A}_k}-\nu)^2}.
\end{align}
In Fig.~\ref{fig5}(b) the total population is shown following a pulsed excitation with the laser frequency tuned to the superradiant mode. The numerical fit shows that most population is trapped into dark states with an effective overall decay constant equal to $\Gamma_0(1-e^{-\lambda^2})$, as predicted in Sec.~\ref{singleex}.

\section{Molecular coherent light sources}
\label{Sec5}

The formalism developed in Sec.~\ref{Sec3} allows us to tackle platforms such as molecular nano-rings illuminated by incoherent light, as recently advanced in Ref.~\cite{holzinger2020nanoscale}. It has been suggested that these might act as natural filters with coherent light as output. The situation is illustrated in Fig.~\ref{fig1}(d): an incoherently pumped (at rate $\eta_p$) central emitter couples to the waveguide-like light modes supported by the ring of surrounding $\mathcal{N}$ emitters. While the treatment in Ref.~\cite{holzinger2020nanoscale} has been restricted to ideal, identical two level systems and strongly relied on numerical evidence, we aim here at providing a deeper analytical understanding and the natural extension to more complex, molecular quantum emitters. Our analysis is based on simplifications brought on by the transition from the bare basis to the collective basis.\\
\indent We will make use of results in Sec.~\ref{Sec3} and notice that the central pump molecule is solely coupled to the symmetric combination of the ring molecules with the Hamiltonian
\begin{align}
\label{laserHam}
\mathcal{H}_p = \omega_p \sigma^\dagger_p \sigma_p + \sqrt{\mathcal{N}}\Omega_p^\lambda(d) \left[\sigma^\dagger_p \mathcal{S} + \mathcal{S}^\dagger \sigma_p\right].
\end{align}
As the symmetric operator creates delocalized excitations over the whole ring, the coupling above benefits from the collective enhancement with $\sqrt{\mathcal{N}}$ multiplying the dipole-dipole exchange rate $\Omega_p^\lambda(d)$ which is dependent on the ring radius $r=d/[2\sin{(2\pi/\mathcal{N})}]$. Notice that the effect of vibrations has already been taken into account by the renormalization of any dipole-dipole coherent and incoherent exchanges with the Huang-Rhys factor (denoted by the index $\lambda$). The effect is mainly detrimental as the coherent coupling between the pump emitter and the waveguide emitters is scaled down both with $\lambda$ and with temperature.\\
The dissipative part of the master equation governing the whole system's evolution includes the usual terms characterizing the decay of the ring molecules, adding to the diagonal decay of the pump molecule and the mutual incoherent coupling between pump and ring molecules of the form
\begin{align}
\mathcal{L}_p[\rho] = \frac{\sqrt{\mathcal{N}}\Gamma_p^\lambda(d)}{2}\Big[ 2\mathcal{S} \rho \sigma_p^\dagger +2\sigma_p \rho \mathcal{S}^\dagger - \{ \mathcal{S}^\dagger \sigma_p + \sigma_p^\dagger \mathcal{S},\rho \}  \Big],
\end{align}
where $\Gamma_p^\lambda(d)$ the incoherent coupling between the pump molecule and each of the ring molecules. In addition, incoherent pump is modeled as an inverted spontaneous emission process: this is in Lindblad form but with a collapse operator $\sigma_p^\dagger$ and rate $\eta_p$
\begin{align}
\mathcal{L}_{\eta_p}[\rho] = \frac{\eta_p}{2}\Big[ 2\sigma_p^\dagger  \rho \sigma_p - \{ \sigma_p \sigma_p^\dagger ,\rho \}  \Big].
\end{align}
\begin{figure}[t]
\includegraphics[width=0.99\columnwidth]{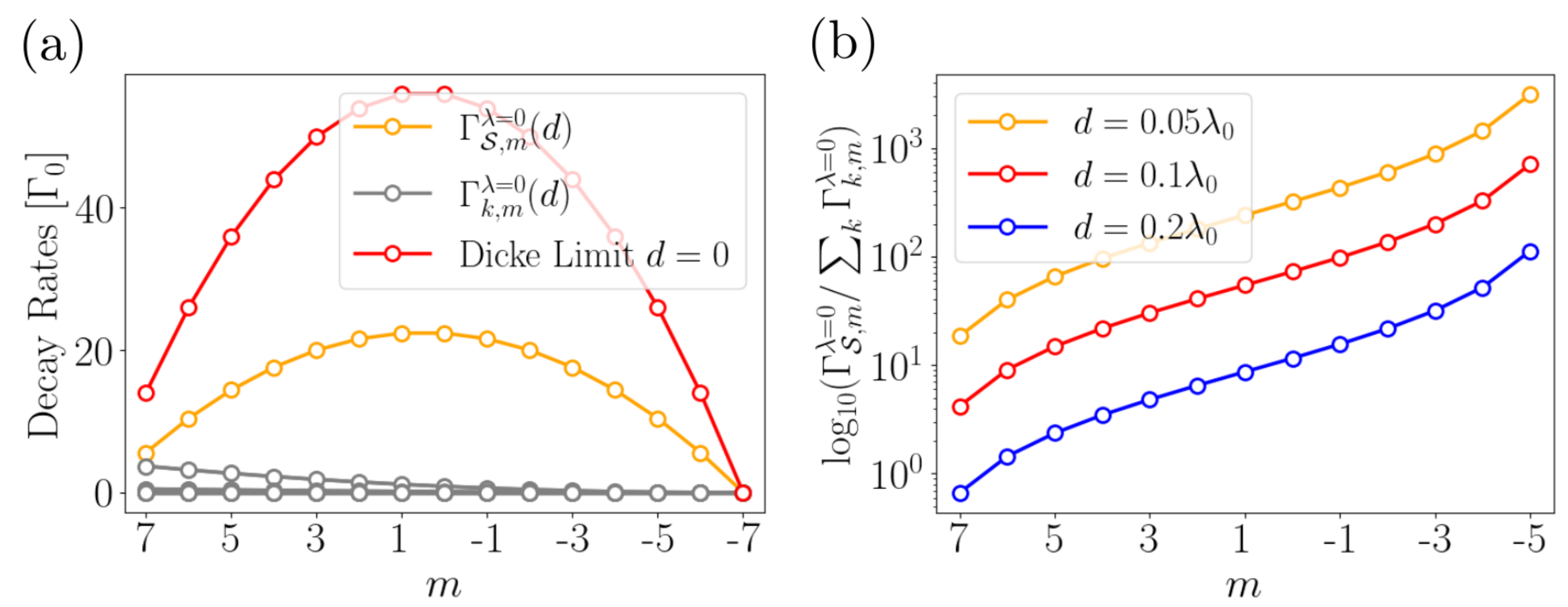}
\caption {(a) State dependent decay rates via symmetric and antisymmetric loss channels for the ring configuration of $\mathcal{N} = 14$ emitters placed in in the $xy$-plane with separation $d=0.1\uplambda_0$ and dipole polarization in the $z$-direction. A comparison with the full Dicke limit is provided. (b) The ratio between the state dependent symmetric decay rate and the sum of the dark decay rates as a function of the inversion quantum number $m$. It can be seen that loss of excitations takes place mainly via the symmetric decay channel even at distances of the order $d=0.2\uplambda_0$.}
\label{fig6}
\end{figure}
\indent To characterize the emission properties of the system, one makes use of both the emitted light intensity $\mathcal{I}_\text{out}$ as well as of the $g^{(2)}$-function at zero time delay. We proceed by using the definitions from Ref.~\cite{holzinger2020nanoscale} in the uncoupled basis before performing our analysis in the alternative collective basis. The intensity in the bare, uncoupled basis is a sum over the following terms $\mathcal{I}_\text{out}=\sum_{jj'}^{\mathcal{N}+1} \Gamma_{jj'}^\lambda \langle \sigma^\dagger_j \sigma_{j'} \rangle$, where now the sum extends to the additional site which is the pump molecule. In the collective basis this can be expressed as
\begin{align}
\label{Iout}
\mathcal{I}_\text{out} &= \Gamma_\mathcal{S}^\lambda (d) \langle \mathcal{S}^\dagger \mathcal{S} \rangle + \sum_{k=1}^{\mathcal{N}-1}\Gamma_k^\lambda (d) \langle \mathcal{A}_k^\dagger \mathcal{A}_k \rangle \\
 &+ 2 \sqrt{\mathcal{N}} \Gamma_p^\lambda(d) \mathrm{Re}\langle \mathcal{S}^\dagger \sigma_p\rangle + \Gamma_\mathrm{0} \langle \sigma^\dagger_p \sigma_p \rangle. \nonumber
\end{align}
For small inter-emitter separation $d \ll \uplambda_0$ the ring contribution can be expressed purely in terms of the symmetric mode as decay into antisymmetric states is negligible $\Gamma_k^\lambda(d) / \Gamma_0 \ll 1$. This can be easily justified by computing the branching of loss rates from a given symmetric state $|\mathcal{N}/2,m \rangle$ into the symmetric manifold and outside of it, into any dark decay channel $k$ by via the Lindbladian in Eq.~\eqref{Lstandard}. One obtains the state dependent decay rates $\Gamma_{k,m}^{\lambda=0}(d)={\alpha_m^{(-)2}}\Gamma_k^{\lambda=0}(d)/(\mathcal{N}-1)$ and the state dependent decay into the symmetric channel $\Gamma_{\mathcal{S},m}^{\lambda=0}(d) ={ \alpha_m^{(-)2}}\Gamma_\mathcal{S}^{\lambda=0}(d)$. In Fig.~\ref{fig6} we plot these rates as a function of the quantum number $m$ as well as the ratio $\Gamma_{\mathcal{S},m}^{\lambda=0} / \sum_k \Gamma_{k,m}^{\lambda=0}$ to show that the restriction of the dynamics to the symmetric subspace is a good approximation for small but still finite distances.\\
\begin{figure}[t]
\includegraphics[width=0.99\columnwidth]{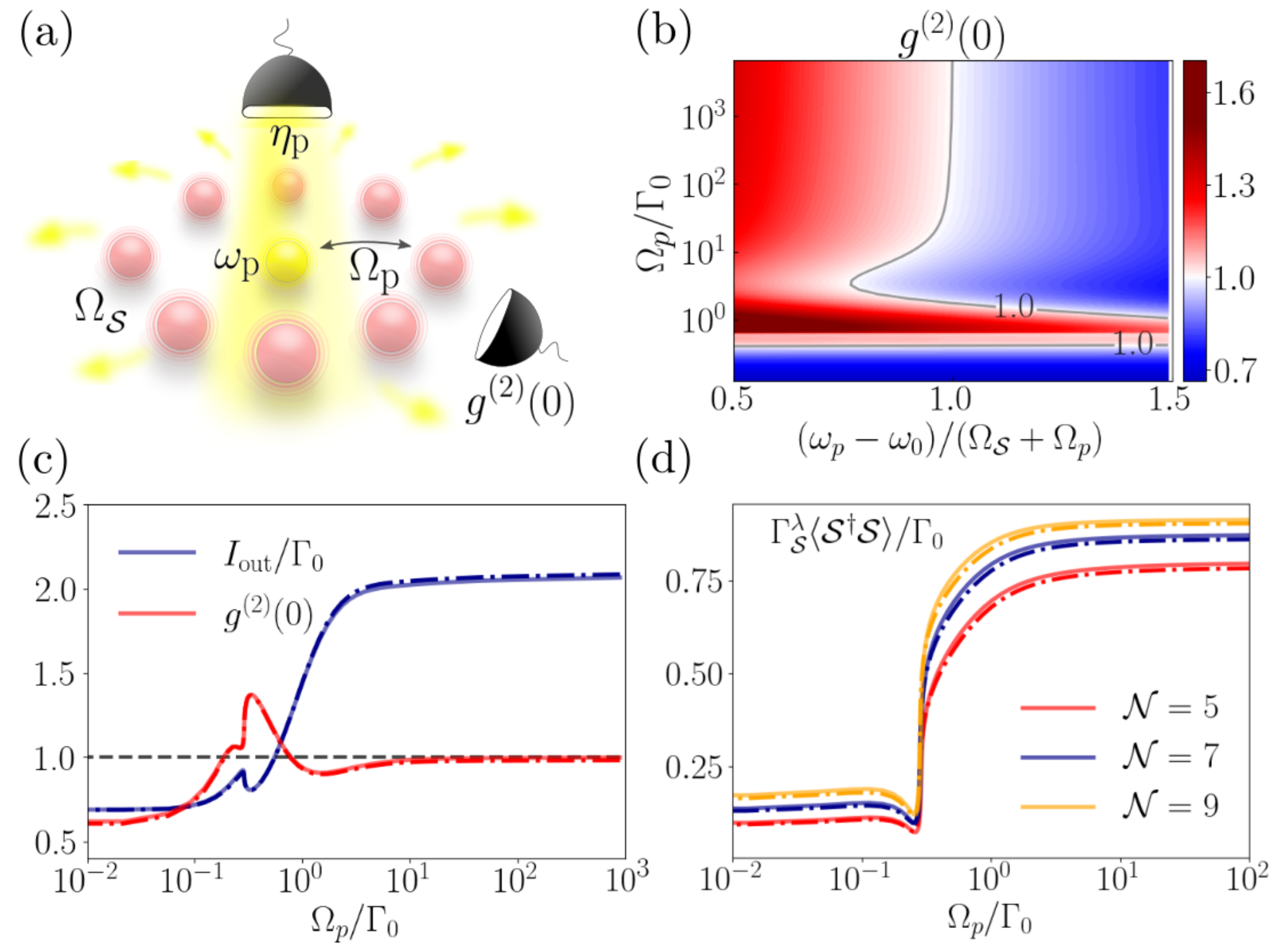}
\caption {(a) Molecular ring acting as a waveguide coupled to a central, incoherently pumped molecule, at electronic transition frequency $\omega_p$ optimally adjusted to fit a waveguide resonance $\omega_0+\Omega_\mathcal{S}^{\lambda=0}+\Omega_p^{\lambda=0}$. (b) The $g^{(2)}(0)$-function in the case of $\mathcal{N}=5$ ring emitter in the absence of vibronic coupling for various tunings $\omega_p$ and coupling strengths $\Omega_p$ for $\eta_p = 3\Gamma_0$. (c) A cut along the optimal resonance frequency $\omega_p$ showing the steady state emission rate alongside the $g^{(2)}(0)$-function for $\eta_p = 3 \Gamma_0,r=0.05\uplambda_0$.
(d) Steady state photon emission for $\eta_p=\Gamma_0$ taking only the symmetric ring contribution in Eq.~\eqref{Iout} into account. A clear threshold for the ring emission emerges at a coupling strength $\Omega_p^{\lambda=0}/\Gamma_0\approx 1$. The dashed-dotted lines represent $\lambda=0.15$ and the continuous lines $\lambda=0$ in all plots.}
\label{fig7}
\end{figure}
\indent Moreover, this approximation is also well justified under weak excitation conditions and with small vibronic couplings. The reason is transparent from Eq.~\eqref{laserHam} which shows that the incoherent pump of the central molecules feeds only the symmetric mode which, in the single excitation regime can only decay back to the ground state, thus not allowing to trap population into robust, antisymmetric states. This is no longer true at higher excitations and in the presence of strong vibronic coupling, where antisymmetric collapse operators can bring population out of the symmetric manifold. In Fig.~\ref{fig7}(d) we illustrate the intensity of emitted light taking only the symmetric ring mode into account, namely $\Gamma_\mathcal{S}^\lambda (d) \langle \mathcal{S}^\dagger \mathcal{S} \rangle$ as a function of the coupling strength with a threshold at $\Omega_p^\lambda(d)\approx \Gamma_0$ after which the emission intensity is sharply increasing. While analytical calculations are possible for a wide range of parameters, the results are cumbersome; we therefore restrict here to the simplified case with $\Gamma_p^\lambda=0$ (for full set of equations see Appendix~\ref{D}):
\begin{align}
\braket{\mathcal{S}^\dagger\mathcal{S}}=\frac{\mathcal{N}\bar{\Gamma}\eta_p{\Omega^\lambda_p}^2}{\Gamma_\mathcal{S}^\lambda(\Gamma_0+\eta_p)\left[\left({\bar{\Gamma}}/{2}\right)^2+{\Omega_\mathcal{S}^\lambda}^2\right]+\bar{\Gamma}^2\mathcal{N}{\Omega_p^\lambda}^2},
\end{align}
where $\bar{\Gamma}=\Gamma_0+\eta_p+\Gamma_\mathcal{S}^\lambda$. The situation is relevant for the ideal geometry chosen in Ref.~\cite{holzinger2020nanoscale} which insured a maximal coherent coupling between the pumped, central emitter while allowing for the mutual dissipative coupling to vanish.\\
\indent In order to characterize statistics of the emitted light, the second order correlation function with zero time delay is used which is defined via the electric field radiated by an ensemble of dipole emitters \cite{gardiner2004quantum}.
Due to the symmetrical ring geometry, the $g^{(2)}$-function in steady state at a detection distance $ |r| \gg \uplambda_0$ in the plane of the ring can be expressed purely in terms of the electronic transition operators as~\cite{holzinger2020nanoscale}
\begin{widetext}
\begin{align}
g^{(2)}(0) = \frac{\sum_{ijkl}^{\mathcal{N}+1} \langle \sigma^\dagger_i \sigma^\dagger_j \sigma_k \sigma_l \rangle }{ \Big( \sum_{ij}^{\mathcal{N}+1} \langle \sigma^\dagger_i \sigma_j \rangle \Big)^2}
= \frac{ 4\mathcal{N}\langle \mathcal{S}^\dagger \mathcal{S} \sigma_p^\dagger \sigma_p \rangle +4\mathcal{N}^{\frac{3}{2}} \mathrm{Re} \langle \mathcal{S}^\dagger \mathcal{S}^\dagger \mathcal{S} \sigma_p\rangle +\mathcal{N}^2 \langle \mathcal{S}^\dagger \mathcal{S}^\dagger \mathcal{S} \mathcal{S}\rangle }
 { \Big(  \mathcal{N}\langle \mathcal{S}^\dagger \mathcal{S} \rangle + 2 \sqrt{\mathcal{N}}\mathrm{Re}\langle \mathcal{S}^\dagger \sigma_p\rangle + \langle \sigma^\dagger_p \sigma_p \rangle \Big)^2}.
\end{align}
\end{widetext}

A second order correlation function equal to unity is used as a figure of merit for coherent light emission and in Fig.~\ref{fig7}(b) it is shown that an optimal resonance frequency $\omega_p = \omega_0+\Omega_\mathcal{S}^{\lambda=0}+\Omega_p^{\lambda=0}$ for the central molecule leads to coherent light emission in particular in the strong coupling regime $\Omega_p^{\lambda=0} \gg \Gamma_0$.
Setting the optimal resonance frequency for the pumped molecule, Fig.~\ref{fig7}(c) shows the total steady state intensity alongside the $g^{(2)}(0)$ as a function of the coupling strength where the sudden increase of intensity stems from the ring contribution as shown in Fig.~\ref{fig7}(d). This sudden increase originates from a coupling strength which attains the same magnitude as the incoherent loss rate into the vaccuum modes $\Gamma_0$ of the pumped molecule. Consequently, in the strong coupling regime the majority of the excitation in the center is coherently transfered to the ring.
\\

\section{Conclusions}
\label{Conclusions}
We have provided a largely analytical approach to the description of light-matter cooperativity in molecular arrays, where subwavelength emitter-emitter separations lead to the occurrence of a strong coherent and incoherent collective response. The effect of molecular vibrations has been incorporated via the Holstein Hamiltonian, that describes vibronic coupling between electronic and nuclear degrees of freedom. In a first step, we have identified analytical scaling laws which characterize phenomena such as super- and subradiance in molecular rings. The ring configuration, as characterized by periodic boundary conditions, allow for the natural extension to mesoscopic systems. For the situation of Dicke superradiance, we find that a collective basis description provides insight into how the superradiant pulse intensity is lost into antisymmetric, dark channels coupled via vibrations. In the low excitation regime, we have analyzed the open system band diagram and found the imprint of the vibrational coupling on both energy and loss rate bands. For molecular dimers, in which case near field couplings are considerably large, we have shown that long-lived bipartite entanglement at the level of electronic degrees of freedom can be produced via dissipative effects such as vibrational relaxation. For incoherently pumped, nanoscale coherent light sources, we have provided analytical results supplementing the results in Ref.~\cite{holzinger2020nanoscale} and an extension to molecular emitters.\\

\textbf{Acknowledgments} -- We acknowledge fruitful discussions with Christian Sommer. We acknowledge financial support from the Max Planck Society and the Deutsche Forschungsgemeinschaft (DFG, German Research Foundation) -- Project-ID 429529648 -- TRR 306 QuCoLiMa
(``Quantum Cooperativity of Light and Matter''). M.~R. acknowledges financial support from the International Max Planck Research School - Physics of Light (IMPRS-PL). R.~H. acknowledges funding from the Austrian Science Fund
(FWF) doctoral college DK-ALM W1259-N27.

\onecolumngrid

\appendix

\section{Vibronic coupling}
\label{AppendixA}

Let us justify the form of the Holstein Hamiltonian in Eq.~\eqref{holstein} by following a first-principle derivation for a single nuclear coordinate $R$ of effective mass $\mu$. We assume that, along the nuclear coordinate, the equilibria for ground (coordinate $R_g$ , state vector $|g\rangle$) and excited (coordinate $R_e$ and state vector $|e\rangle$) electronic orbitals are different.
Assuming equilibrium positions $R_g$ and $R_e$ for the potential surfaces of electronic ground and excited states, one can write the total molecular Hamiltonian describing both electronic and vibrational dynamics as
\begin{align}
\label{molham}
\mathcal{H}_\text{mol}=\left[\omega_0+\frac{\hat{P}^2}{2\mu}+\frac{1}{2}\mu\nu^2\left(\hat{R}-R_e\right)^2\right]\sigma^\dagger\sigma+ \left[\frac{\hat{P}^2}{2\mu}+\frac{1}{2}\mu\nu^2\left(\hat{R}-R_g\right)^2\right]\sigma\sigma^\dagger,
\end{align}
where $\mu$ is the reduced mass of the vibrational mode. The kinetic and potential energies are written in terms of the position $\hat{Q}$ and momentum operator $\hat{P}$ describing the nuclear coordinate under consideration, with commutation $[\hat{Q},\hat{P}]=i$. Introducing oscillations around the equilibria $\hat{Q}=\hat{R}-R_{\text{g}}$ and subsequently $\hat{R}-R_{\text{e}}=\hat{Q}+R_{\text{g}}-R_{\text{e}}=:\hat{Q}-R_{\text{ge}}$ we obtain
\begin{equation}
\mathcal{H}_\text{mol}=\frac{\hat{P}^2}{2\mu}+\frac{1}{2}\mu\nu^2 \hat{Q}^2 +\omega_0\sigma^\dagger\sigma-\mu\nu^2 \hat{Q} R_{\text{ge}}\sigma^\dagger\sigma+\frac{1}{2}\mu\nu^2 R_{\text{ge}}^2 \sigma^\dagger\sigma.
\end{equation}
 We can now rewrite the momentum and position operators in terms of bosonic operators $\hat{Q}=q_{\text{zpm}}(b^\dagger+b)$, $\hat{P}=ip_{\text{zpm}}(b^\dagger-b)$. The bosonic operators satisfy the usual commutation relation $[{b},{b^\dagger}]=1$ and the zero-point motion displacement and momentum are defined as $q_{\text{zpm}}=1/\sqrt{2\mu\nu}$ and $p_{\text{zpm}}=\sqrt{\mu\nu/2}$. Reexpressing the terms above yields the Holstein Hamiltonian \cite{holstein1959study}
\begin{align}
\label{holsteinhamiltonian}
\mathcal{H}_\text{mol}=(\omega_0+\lambda^2\nu){\sigma}^\dagger{\sigma}+\nu {b}^\dagger{b}-\lambda\nu({b}^\dagger+{b}){\sigma}^\dagger{\sigma}.
\end{align}
The dimensionless vibronic coupling strength $\lambda$ is given by $\lambda=\mu\nu R_{\text{ge}}q_{\text{zpm}}$ ($\lambda^2$ is called the Huang-Rhys factor and is typically on the order of $\sim 0.01-1$).
\subsection{The polaron transformation}
The Holstein Hamiltonian can be diagonalized via a level-dependent polaron transformation $\mathcal{U}^\dagger=\ket{g}\bra{g}+\mathcal{D}^\dagger\ket{e}\bra{e}$ with the standard displacement operator $\mathcal{D}=e^{-i\sqrt{2}\lambda p}=e^{\lambda(b^\dagger-b)}$. In the polaron-displaced basis, the Holstein Hamiltonian becomes $\tilde{\mathcal{H}}_\text{mol}=\mathcal{U}^\dagger \mathcal{H}_\text{mol}\mathcal{U}=\omega_0{\sigma}^\dagger{\sigma}+\nu {b}^\dagger{b}$ and has simple eigenvectors $\ket{g;n}$ and $\ket{e;n}$. The eigenvectors in the bare, original basis can be found by inverting the polaron transformation $\ket{g;n}$ and $\mathcal{D}\ket{e;n}$. The polaron-transformed probe Hamiltonian is then expressed as $\tilde{\mathcal{H}}_\ell=i\eta (\sigma^\dagger \mathcal{D}^\dagger e^{-i\omega_\ell t}-\sigma \mathcal{D} e^{i\omega_\ell t})$. One can now look for selection rules applying to processes such as stimulated emission and absorption induced by the external optical drive. To this end, we focus on absorption (as emission is similar) by assuming an initial state $\ket{g;0}$ in the displaced basis and asking for the probability of exciting the system to state $\ket{e;n}$. This is easily computed to lead to
\begin{equation}
\label{poissonian}
P_\text{abs}(n)=|\bra{e;n}\sigma^\dagger\mathcal{D}^\dagger\ket{g;0}|^2=e^{-\lambda^2}\frac{\lambda^{2n}}{n!},
\end{equation}
which is the expected Poissonian distribution leading to the Franck-Condon principle for molecular transitions. For dissipative radiative processes, we notice that the Lindblad collapse operator is also transformed to the polaron one $\sigma \mathcal{D}$ such that spontaneous emission follows the same Poissionian distribution in taking the electronic state from $\ket{e;0}$ to $\ket{g;n}$.\\
\subsection{Thermal averaging of vibrational effects}
Assuming a thermal state for the vibrational modes we are going to calculate the trace of a single vibrational displacement operator $\mathcal{D}^\dagger = e^{-\lambda^2/2}e^{-\lambda b^\dagger} e^{\lambda b}$:
\begin{align}
\langle \mathcal{D}^\dagger \rangle _T&= \mathrm{Tr}[\mathcal{D}^\dagger \rho_{\text{th}}] = e^{-\lambda^2/2} \mathrm{Tr}[e^{-\lambda b^\dagger} e^{\lambda b} \rho_{\text{th}}] = \\
&= e^{-\lambda^2/2}\sum_{n=0}^\infty e^{-\beta\nu n}(1-e^{-\beta\nu}) \braket{n|\sum_{m,l}\frac{(-\lambda^m)\lambda^l}{m!l!}(b^\dagger)^m b^l|n} \nonumber \\
&= e^{-\lambda^2/2}\sum_{n=0}^\infty e^{-\beta\nu n}(1-e^{-\beta\nu}) \sum_{m=0}^n \frac{(-\lambda^2)^m}{m!}\binom{n}{m} \nonumber \\
&=e^{-\lambda^2/2}(1-e^{-\beta\nu})\sum_{m=0}^\infty \frac{(-\lambda^2)^m}{m!} \sum_{n=m}^\infty e^{-\beta\nu n} \binom{n}{m} \nonumber ,
\end{align}
where we made use of the sum identity $\sum_{i=k}^n\sum_{j=k}^i a_{i,j}=\sum_{j=k}^n\sum_{i=j}^n a_{i,j}$ in the last step. Additionally making use of the binomial identity $\sum_{n=k}^\infty\binom{n}{k}y^n=\frac{y^k}{(1-y)^{k+1}}$ one readily obtains
\begin{align}
\langle \mathcal{D}^\dagger \rangle _T=e^{-\lambda^2/2}\sum_{m=0}^\infty\frac{(-\lambda^2)^m}{m!}\frac{e^{-\beta\nu m }}{(1-e^{-\beta\nu})^m}=e^{-\lambda^2/2(1+2\bar{n})}=e^{-\frac{\lambda^2}{2}\coth\left(\frac{\hbar\nu}{2k_B T}\right)}.
\end{align}

\section{Vacuum mediated coherent and incoherent coupling rates }
\label{AppendixB}
The vacuum mediated dipole-dipole interactions for an electronic transition at wavelength $\uplambda_0$ (corresponding wave vector $k=2\pi/\uplambda_0$) between an identical pair of emitters separated by $r_{ij}$ is
\begin{equation}
\Omega_{ij}=\frac{3}{4}\Gamma_0 \left[(1-3\cos^2\theta)\left(\frac{\sin (kr_{ij})}{(kr_{ij})^2}+\frac{\cos (kr_{ij})}{ (kr_{ij})^3}\right)-\sin^2\theta\frac{\cos (kr_{ij})}{(kr_{ij})}\right].
\end{equation}
The quantity $\theta$ is the angle between the dipole moment $\mathbf{d}$ and the vector $\mathbf{r}_{ij}$. The associated collective decay is quantified by the following mutual decay rates
\begin{equation}
\Gamma_{ij}=\frac{3}{2}\Gamma_0 \left[(1-3\cos^2\theta)\left(\frac{\cos (kr_{ij})}{(kr_{ij})^2}-\frac{\sin (kr_{ij})}{ (kr_{ij})^3}\right)+\sin^2\theta\frac{\sin (kr_{ij})}{(kr_{ij})}\right].
\end{equation}

\section{Vibrationally mediated energy transfer rates in the collective basis}
\label{C}

The Holstein Hamiltonian rewritten in a collective basis both for the electronic as well as the vibrational degrees of freedom has the following form
\begin{align}
\mathcal{H}^\text{dim} = \omega_\mathcal{S}(Q_+)\mathcal{S}^\dagger \mathcal{S} + \omega_\mathcal{A}(Q_+)\mathcal{A}^\dagger \mathcal{A}
- \frac{\lambda \nu}{\sqrt{2}} Q_-(\mathcal{S}^\dagger \mathcal{A} + \mathcal{A}^\dagger \mathcal{S}) + \nu \sum_{k= \pm} b^\dagger_k b_k ,
\end{align}
where the energies of the collective state frequencies depend on the symmetric vibrational coordinate $\omega_\mathcal{S}(Q_+)=\omega_0+\lambda^2\nu+\Omega-\lambda\nu Q_+/\sqrt{2}$ and $\omega_\mathcal{A}(Q_+)=\omega_0+\lambda^2\nu-\Omega-\lambda\nu Q_+/\sqrt{2}$. We note that the $Q_+$-dependent shifts can be removed by the collective polaron transforms $\mathcal{U}_\mathcal{S}=e^{i\lambda P_+\mathcal{S}^\dagger\mathcal{S}/\sqrt{2}}$ and $\mathcal{U}_\mathcal{A}=e^{i\lambda P_+\mathcal{A}^\dagger\mathcal{A}/\sqrt{2}}$ which transform the symmetric nuclear coordinates as
\begin{subequations}
\begin{align}
\mathcal{U}_\mathcal{S} Q_+ \mathcal{U}_\mathcal{S}^\dagger &= Q_++\sqrt{2}\lambda\mathcal{S}^\dagger\mathcal{S},\\
\mathcal{U}_\mathcal{A} Q_+  \mathcal{U}_\mathcal{A}^\dagger &= Q_++\sqrt{2}\lambda\mathcal{A}^\dagger\mathcal{A},
\end{align}
\end{subequations}
and lead to a renormalization of the state energies $\tilde{\omega}_\mathcal{A}=\omega_0+\lambda^2\nu/2+\Omega$ and $\tilde{\omega}_\mathcal{S}=\omega_0+\lambda^2\nu/2-\Omega$. The equation of motion for the operators are given by
\begin{subequations}
\begin{align}
\dot{\mathcal{S}} &= -\Big[i \tilde{\omega}_\calS+ \frac{\Gamma_\calS}{2}\Big] \mathcal{S} +\frac{i\lambda \nu}{\sqrt{2}} Q_- \mathcal{A} + \sqrt{\Gamma_\calS} \mathcal{S}_\mathrm{in}, \\
\dot{\mathcal{A}} &= -\Big[i\tilde{\omega}_\calA + \frac{\Gamma_\calA}{2}\Big] \mathcal{A} +\frac{i\lambda \nu}{\sqrt{2}} Q_- \mathcal{S} + \sqrt{\Gamma_\calA} \mathcal{A}_\mathrm{in},
\end{align}
\end{subequations}
where $\mathcal{S}_\mathrm{in} = (\sigma_\mathrm{1,in}+\sigma_\mathrm{2,in})/\sqrt{2}$ and $\mathcal{A}_\mathrm{in} = (\sigma_\mathrm{1,in}-\sigma_\mathrm{2,in})/\sqrt{2}$ are the collective noise terms which we will neglect from now on as they do not contribute to the transfer process.

To calculate the transfer rate from the symmetric state to the antisymmetric state we assume some initial population in the symmetric state and no population in the antisymmetric state, additionally we assume that the symmetric state decays independently and formally integrate
\begin{subequations}
\begin{align}
\mathcal{S}(t) &= S(0)e^{-(i\tilde{\omega}_\calS+\Gamma_\calS/2)t}, \\
\mathcal{A}(t) &= A(0)e^{-i(\tilde{\omega}_\calA+\Gamma_\calA/2)t} +\frac{i\lambda \nu}{\sqrt{2}}\int^t_0 dt'e^{-i(\tilde{\omega}_\calA+\Gamma_\calA/2)(t-t')}Q_-(t')\mathcal{S}(t'),
\end{align}
\end{subequations}
and for the expectation value of the populations we get
\begin{subequations}
\begin{align}
\dot{\langle \mathcal{S}^\dagger \mathcal{S} \rangle} &= -\Gamma_\calS{\langle \mathcal{S}^\dagger \mathcal{S} \rangle} -\sqrt{2} \lambda \nu\, \mathrm{Im}\langle \mathcal{S}^\dagger \mathcal{A} Q_-\rangle ,\\
\dot{\langle \mathcal{A}^\dagger \mathcal{A} \rangle} &= -\Gamma_\calA{\langle \mathcal{A}^\dagger \mathcal{A} \rangle} -\sqrt{2} \lambda \nu\, \mathrm{Im}\langle \mathcal{A}^\dagger \mathcal{S} Q_-\rangle.
\end{align}
\end{subequations}

\begin{figure}[t]
    \begin{center}
      \includegraphics[width=0.6\textwidth]{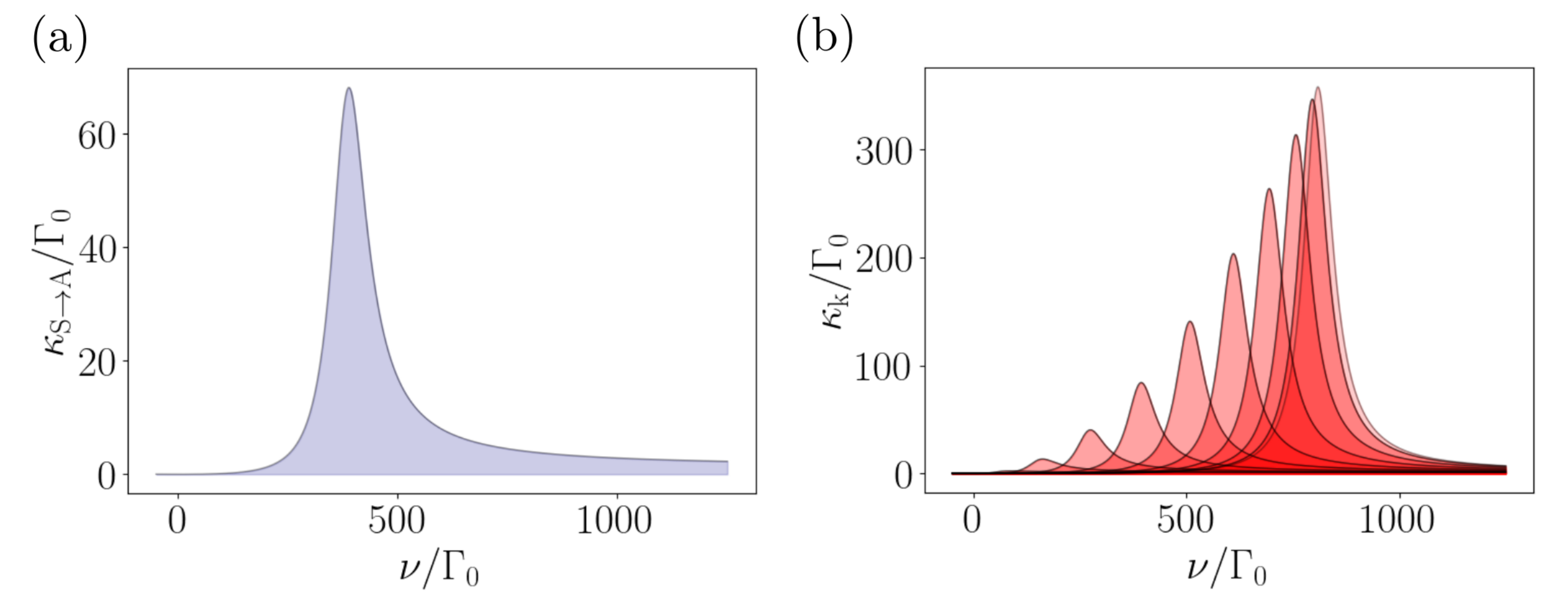}
     \end{center}
    \caption{(a) Transfer rate from the symmetric to the antisymmetric state for a molecular dimer as a function of the vibronic frequency. The maximum transfer occurs at the resonance $\nu = \Omega_S^\lambda-\Omega_A^\lambda$. (b) Transfer rates for a molecular ring of $\mathcal{N}=20$ molecules from the symmetric state with mode number $k=0$ to the $\mathcal{N}-1$ dark states. Resonances occur at $\nu = \Omega_S^\lambda-\Omega_{k}^\lambda$ for $k = 1,...,\lceil (\mathcal{N}-1)/2 \rceil$ and the range of vibrational frequencies at which transfer to the dark state manifold occur is increasing with $\mathcal{N}$ as well as the linewidth of the vibrational resonance $\Gamma_\nu$. Parameters are $d=0.025 \uplambda_0, \ \Gamma_\nu = 100\Gamma_0, \ \lambda = 0.15$ and the dipole polarization is chosen perpendicular to the ring plane.}%
    \label{figA3}%
\end{figure}

Therefore $-\sqrt{2} \lambda \nu \,\mathrm{Im}\langle \mathcal{A}^\dagger \mathcal{S} Q_-\rangle$ will be responsible for population transfer from the symmetric to the antisymmetric state at a rate $\kappa_{\mathcal{S} \rightarrow \mathcal{A}}$.

\begin{align}
-\sqrt{2} \lambda \nu \langle \mathcal{A}^\dagger \mathcal{S} Q_-\rangle &= -i\lambda^2 \nu^2 \int_0^t dt' e^{-\epsilon_\calA(t-t')} \langle Q_-(t')Q_-(t)\rangle \langle \mathcal{S}^\dagger(0) \mathcal{S}(0)\rangle e^{-\epsilon_\calS t'} e^{-\epsilon^*_\calS t} \nonumber \\
 &= -i \lambda^2 \nu^2 \langle \mathcal{S}^\dagger(0) \mathcal{S}(0)\rangle \frac{e^{-\Gamma_\calS t} - e^{-((\Gamma_\nu+\Gamma_\calS-\Gamma_\calA)/2+i(\omega_\calS-\omega_\calA-\nu))t}}{(\Gamma_\nu+\Gamma_\calA-\Gamma_\calS)/2 + i(\omega_\calS-\omega_\calA-\nu)},
\end{align}
where we used the fact, that the expectation values for $\mathcal{S}$ and $Q_-$ factorize and defined $\epsilon_\calA = -(\Gamma_\calA/2-i\tilde{\omega}_\calA)$ and $\epsilon_\calS = -(\Gamma_\calS/2-i\tilde{\omega}_\calS)$. The correlations for $Q_-$ are evaluated assuming free evolution of the vibrations (to lowest order) and  zero temperature for the vibrational modes:
\begin{equation}
\langle Q_-(t') Q_-(t) \rangle = \frac{1}{2}\Big(\langle b_1(t')b_1^\dagger(t)\rangle+\langle b_2(t')b_2^\dagger(t)\rangle \Big) =e^{-(\Gamma_\nu/2-i\nu)(t-t')}.
\end{equation}
In the case of a fast vibrational relaxation rate $\Gamma_\nu \gg \Gamma_\calS,\Gamma_\calA$ the transfer rate can be written as:
{\begin{equation}
 \kappa_{\mathcal{S} \rightarrow \mathcal{A}} = \frac{\lambda^2 \nu^2}{2} \frac{\Gamma_\nu +\Gamma_\calA-\Gamma_\calS}{\frac{(\Gamma_\nu+\Gamma_\calA-\Gamma_\calS)^2}{4}+ (\omega_\calS-\omega_\calA-\nu)^2}.
\end{equation}}
The transfer rate from the antisymmetric to the symmetric state can be calculated similarly, assuming initial population in the antisymmetric state:

{\begin{equation}
 \kappa_{\mathcal{A} \rightarrow \mathcal{S}} = \frac{\lambda^2 \nu^2}{2} \frac{\Gamma_\nu +\Gamma_\calS-\Gamma_\calA}{\frac{(\Gamma_\nu+\Gamma_\calS-\Gamma_\calA)^2}{4} + (\omega_\calA-\omega_\calS-\nu)^2}.
\end{equation}}

\subsection*{Generalization to $\mathcal{N}$  molecules}

The generalization to an arbitrary number of molecules is straightforward by first writing the full Hamiltonian in a collective basis for both the electronic as well as the vibrational modes.

\begin{align}
\mathcal{H}_\mathrm{coll} &= \omega_\mathcal{S}(Q_\mathcal{N})\mathcal{S}^\dagger \mathcal{S} + \sum_{k=1}^{\mathcal{N}-1}\omega_k(Q_\mathcal{N})\mathcal{A}_k^\dagger \mathcal{A}_k - \frac{\lambda\nu}{\sqrt{\mathcal{N}}} \sum_{k=1}^{\mathcal{N}-1} (Q_k \mathcal{S}^\dagger \mathcal{A}_k + h.c.) - \frac{\lambda\nu}{\sqrt{\mathcal{N}}} \sum_{k\neq k'}^{\mathcal{N}-1}(Q_{k-k'}\mathcal{A}_k^\dagger \mathcal{A}_{k'} +h.c.)+\nu \sum_{k=1}^{\mathcal{N}} b^\dagger_k b_k,
  \end{align}
where the energies of the collective states are shifted by the contribution of the symmetric vibrational mode $\omega_k(Q_\mathcal{N})=\omega_0+\lambda^2\nu+\Omega_k-\lambda\nu Q_\mathcal{N}/\sqrt{\mathcal{N}}$ for $k=1,\hdots,\mathcal{N}$.
Similarly to the dimer case, the $Q_\mathcal{N}$-dependent energy shifts can be removed by the collective polaron transformation $\prod_{k=1}^\mathcal{N}\mathcal{U}_{\mathcal{A}_k} = \prod_{k=1}^\mathcal{N} e^{i\lambda P_\mathcal{N} \mathcal{A}_k^\dagger \mathcal{A}_k /\sqrt{\mathcal{N}}}$ which leads to a renormalization of the collective state energies as $\tilde{\omega}_k=\omega_0+\lambda^2\nu/2+\Omega_k$. The crucial term is however the coupling between the symmetric states and the dark state manifold.
Similar to the molecular dimer case we assume initial population in the symmetric state and solve the Heisenberg equations of motion neglecting the noise terms

    \begin{subequations}
        \begin{align}
            \dot{\mathcal{S}} &= -i \tilde{\omega}_\mathcal{S} \mathcal{S} - \frac{\Gamma_\mathcal{S}}{2}\mathcal{S} +\frac{i\lambda \nu}{\sqrt{\mathcal{N}}}\sum_{k=1}^{\mathcal{N}-1} Q_k \mathcal{A}_k, \\
          \dot{\mathcal{A}_k} &= -i \tilde{\omega}_k \mathcal{A}_k - \frac{\Gamma_k}{2} \mathcal{A}_k  +\frac{i\lambda \nu}{\sqrt{\mathcal{N}}}Q_k^\dagger \mathcal{S} +\frac{i\lambda \nu}{\sqrt{\mathcal{N}}} \sum_{k'\neq k}^{\mathcal{N}-1} Q_{k-k'} \mathcal{A}_{k'}.
        \end{align}
        \end{subequations}
After solving for the population $\langle \mathcal{S}^\dagger \mathcal{S}\rangle (t)$ and tracing out the vibrational modes one finds transfer rates $\kappa_{\mathcal{S}\rightarrow \mathcal{A}_k}$ between the symmetric state and the dark state manifold:

{\begin{equation}
 \kappa_{\mathcal{S} \rightarrow \mathcal{A}_k} = \frac{\lambda^2 \nu^2}{2} \frac{\Gamma_\nu +\Gamma_{k}-\Gamma_\mathcal{S} }{\frac{(\Gamma_\nu+\Gamma_{k}-\Gamma_\mathcal{S} )^2}{4}+ (\Omega_\mathcal{S} -\Omega_k-\nu)^2}.
\end{equation}}

\section{Equations of motion for coherent light source in symmetric subspace}
\label{D}

In the low-excitation limit for the ring $\braket{\mathcal{S}^z}\approx -\mathcal{N}/2$, a closed set of equations can be obtained describing the interactions between the central pump molecule and the ring (in the symmetric subspace):
\begin{subequations}
\begin{align}
\frac{d}{dt}\braket{\sigma_p^\dagger\sigma_p}&=-(\Gamma_0+\eta_p)\braket{\sigma_p^\dagger\sigma_p}+\eta_p-2\sqrt{\mathcal{N}}\Omega_p^\lambda\,\mathrm{Im}\braket{\calS^\dagger\sigma_p}-\Gamma_p^\lambda\,\mathrm{Re}\braket{\calS^\dagger\sigma_p} ,\\
\frac{d}{dt}\braket{\calS^\dagger\calS}&=-\Gamma_\calS^\lambda\braket{\calS^\dagger\calS}+2\sqrt{\mathcal{N}}\Omega_p^\lambda\,\mathrm{Im}\braket{\calS^\dagger\sigma_p}-\Gamma_p^\lambda\,\mathrm{Re}\braket{\calS^\dagger\sigma_p} ,\\
\frac{d}{dt}\braket{\calS^\dagger\sigma_p}&=-\left(\frac{\Gamma_0+\eta_p+\Gamma_\calS^\lambda}{2}\right)\braket{\calS^\dagger\sigma_p}-i\Omega_\calS^\lambda\braket{\calS^\dagger\sigma_p}+i\sqrt{\mathcal{N}}\Omega_p^\lambda(\braket{\sigma_p^\dagger\sigma_p}-\braket{\calS^\dagger\calS})-\frac{\Gamma_p^\lambda}{2}(\braket{\calS^\dagger\calS}+\braket{\sigma_p^\dagger\sigma_p}).
\end{align}
\end{subequations}
The solutions presented in the main text are then obtained by assuming steady state, which corresponds to setting the time derivatives to zero.

\end{document}